\documentclass{emulateapj}
\usepackage{graphicx,ifthen,url}
\usepackage{apjfonts}
\usepackage{longtable}
\usepackage{ulem}
\usepackage{natbib}
\usepackage{soul}
\usepackage{verbatim}
\newcommand{\Ha}{{\rm\,H$\alpha$}}

\newcommand{\Nii}{{\rm\,[N\,{\sc ii}]}}
\newcommand{\Oii}{{\rm\,[O\,{\sc ii}]}}
\newcommand{\Oiii}{{\rm\,[O\,{\sc iii}]}}
\newcommand{\Sii}{{\rm\,[S\,{\sc ii}]}}
\newcommand{\Siii}{{\rm\,[S\,{\sc iii}]}}

\newcommand{\hi}{H\,{\sc i}\rm}
\newcommand{\hii}{H\,{\sc ii}\rm}

\newcommand{\nii}{[N\,{\sc ii}]}

\newcommand{\oiii}{[O\,{\sc iii}]}

\newcommand{\sii}{[S\,{\sc ii}]}

\newcommand{\lin}{$\,\lambda$}
\newcommand{\llin}{$\,\lambda\lambda$}

\newcommand{\oh}{12\,+\,log(O/H)}

\newcommand{\vs}{vs.}

\shorttitle{Testing for azimuthal abundance gradients in M101}
\shortauthors{Li et al.}

\begin{document} 

\title{Testing for azimuthal abundance gradients in M101}
\author{Yanxia Li$^1$, Fabio Bresolin$^1$ and Robert C. Kennicutt, Jr.$^2$}
\affil{$^1$Institute for Astronomy, 2680 Woodlawn Dr., Honolulu, HI 96822, USA \\
$^2$Institute of Astronomy, University of Cambridge, Madingley Road, Cambridge CB3 0HA, UK}



\begin{abstract}
New optical spectra of 28 \hii\ regions in the M101 disk have been obtained, yielding 10 new detections of the \Oiii$\lambda$4363 auroral line. The oxygen abundance gradient measured from these data, combined with previous observations, displays a local scatter of 0.15$\,\pm\,$0.03 dex along an arc in the west side of the galaxy, compared with a smaller scatter of 0.08$\,\pm\,$0.01 dex in the rest of the disk.
One of the \hii\ regions in our sample (H27) has a significantly lower oxygen abundance than  surrounding nebulae at a similar galactocentric distance, while an additional, relatively nearby one (H128) was already known to have a high oxygen abundance for its position in the galaxy. These results represent marginal evidence for the existence of moderate deviations from chemical abundance homogeneity in the interstellar medium of M101. 
Using a variety of strong-line abundance indicators, we find no evidence for significant large-scale azimuthal variations of the oxygen abundance across the whole disk  of the galaxy. 

\end{abstract}





\section{Introduction}


Radial gradients of metallicity in spiral galaxies have been known for a long time, with typical values of 0.03$-$0.10 dex kpc$^{-1}$. The study of giant extragalactic  \hii\ regions offered the first evidence for the presence  of oxygen abundance gradients in galaxies (\citealt{Searle:1971})
 and has been extensively used  to probe the stellar populations and the chemical composition of nearby star-forming galaxies (\citealt{Vila-Costas:1992,Zaritsky:1994,van-Zee:1998a,Bresolin:1999}), placing strong constraints on galactic chemical evolution models  (\citealt{Chiappini:2001,Fu:2009}). 
In contrast, surprisingly little is known about the possible presence of azimuthal asymmetries in the abundance distribution in spiral disks. From a theoretical point of view, a virtually uniform distribution,  with negligible scatter, could be expected along the azimuthal direction, as a result of  relatively fast mixing processes 
(except near corotation, \citealt{Scarano:2013}), with timescales on the order of $\sim$100~Myr,  in the turbulent interstellar medium (\citealt{Roy:1995, Yang:2012}).

Deviations of the {\it radial} abundance gradient from a simple exponential  seem to be well-established from studies of Cepheids (e.g.~\citealt{Luck:2003}) and open clusters (\citealt{Twarog:1997, Lepine:2011}), showing a discontinuity at a galactocentric distance of $\sim$8.5~kpc and a flattening beyond that radius, possibly caused by the barrier effect of corotation, which isolates the inner and outer regions of the disk one
 from the other, due to opposite directions of gas flow (\citealt{Lepine:2011}). Similar features can be observed in the \hii\ region abundance distribution of nearby spiral galaxies (\citealt{Bresolin:2009, Bresolin:2012, Scarano:2011}).
Investigations of Cepheids (\citealt{Pedicelli:2009}) and \hii\ regions (\citealt{Balser:2011}) in the Milky Way have provided indications for the presence, at least near the corotation radius, of {\it azimuthal} gradients ($\sim$0.05~dex kpc$^{-1}$) and chemical inhomogeneities, which can be attributed to the spiral arm structure and the consequent non-uniformity of the spatial distribution of  gas and star formation in the disk of the Galaxy (\citealt{Lepine:2011}). These results stress the importance of including the azimuthal coordinate, in addition to the radial one, in modeling the chemical evolution of the Milky Way.

The situation in other spiral galaxies is less clear, and suffers from poor statistics (small number of \hii\ regions studied in a given galaxy) and measurement uncertainties.
In the galaxy M33, \citet{Rosolowsky:2008} measured a substantial intrinsic dispersion of 0.11 dex (in addition to the dispersion due to observing errors)
 in the \hii\ region oxygen abundance at constant radius.
However, \citet{Bresolin:2011} showed that this is due to poor detections of the \Oiii$\lambda4363$ auroral line used to derive nebular electron temperatures, and that 
a much smaller scatter of $\sim$0.06 dex, consistent with the measurement uncertainties, is obtained  from the best-quality data.
In a wide-field integral field spectroscopy study of NGC~628, \citet{Rosales-Ortega:2011} found that the radial metallicity gradient  varies slightly for different  quadrants, although the differences are comparable to the uncertainties introduced by  the nebular abundance diagnostics and can be related instead to systematic variations of the ionization parameter. The narrow-band imaging across the disk of the same galaxy  by \citet{Cedres:2012} suggests that the detection of chemical inhomogeneities may depend on the choice of strong-line  diagnostics used to derive the oxygen abundances. 
Recently, a large degree of inhomogeneity on relatively small spatial scales ($\sim$0.5~kpc) has been claimed by \citet{Sanders:2012}  for  the \hii\ region oxygen abundances in M31. This result is not confirmed by other investigations in the same galaxy, including the recent one by \citet{Zurita:2012}, and it is in general still unclear what role the sample selection and choice of chemical abundance diagnostics play in these
results. 

In an attempt to provide new observational constraints on the presence of chemical inhomogeneities and azimuthal gradients in nearby spiral galaxies, we turned to M101 (NGC~5457), a nearby \citep[$D=6.85$ Mpc,][]{Freedman:2001}, nearly face-on, grand-design spiral galaxy, that has been widely studied as a prototype system for the investigation of radial gradients of element abundances (see \citealt{Kennicutt:2003} and \citealt{Bresolin:2007} for a review and references). There are more than $10^3$ cataloged \hii\ regions in its  disk \citep{Hodge:1990}. \citet{Kennicutt:1996}  suggested the presence of a possible azimuthal asymmetry in the oxygen abundance distribution
between the southeast (SE) and northwest (NW) regions of M101, based on strong-line abundance diagnostics. 
However, this result  is subject to the uncertainties and discrepancies concerning the calibrations of these diagnostics (\citealt{Bresolin:2007, Kewley:2008}). 
\citet{Kennicutt:2003}  improved on previous studies using {\it direct} oxygen measurements for 20 \hii\ regions, which rely on the determination of the electron temperature from auroral-to-nebular line ratios, such as \Oiii$\lambda$4363/$\lambda5007$. These authors found a $\sim$0.2 dex metallicity spread for  \hii\ regions located at a similar galactocentric distance in the southwestern section of the galaxy. However, limited by the size of their \hii\ region sample, they could not confirm the large-scale azimuthal asymmetry between the SE and NW, suggested by \citet{Kennicutt:1996}. 

In this paper, we present new spectroscopic observations of 28  \hii\ regions in M101,  and are able to derive  direct oxygen abundances for a subset of 10. This new dataset 
 allows us to address the possibility of a non-axisymmetric distribution of the nebular oxygen abundance in this galaxy. In addition,
because M101 is known to be currently experiencing an infall of high-velocity gas, as shown by \hi\ maps (\citealt{Sancisi:2008}), and has likely been recently subjected to  interaction events, as shown by the lopsidedness of its disk, the peculiar spiral structure and faint tidal structures (\citealt{Waller:1997,Mihos:2012}),  it is well suited to verify to what extent  these events can affect the distribution of metals in a spiral disk.

This paper is organized as follows. In Section~\ref{s:observations} we describe the observations and data reduction procedures. In Section~\ref{s:radial} we present and analyze the radial gradients, and the local azimuthal variations of oxygen abundance, using  both the direct method (\Oiii$\lambda$4363-based) and strong-line diagnostics. We test the asymmetry of the oxygen abundance distribution between the east and west sections of M101 in Section~\ref{s:halves}, and present our conclusions in Section~\ref{s:conclusion}.




\section{Observations}\label{s:observations}
 
\begin{figure*}   
  \centering
    \includegraphics{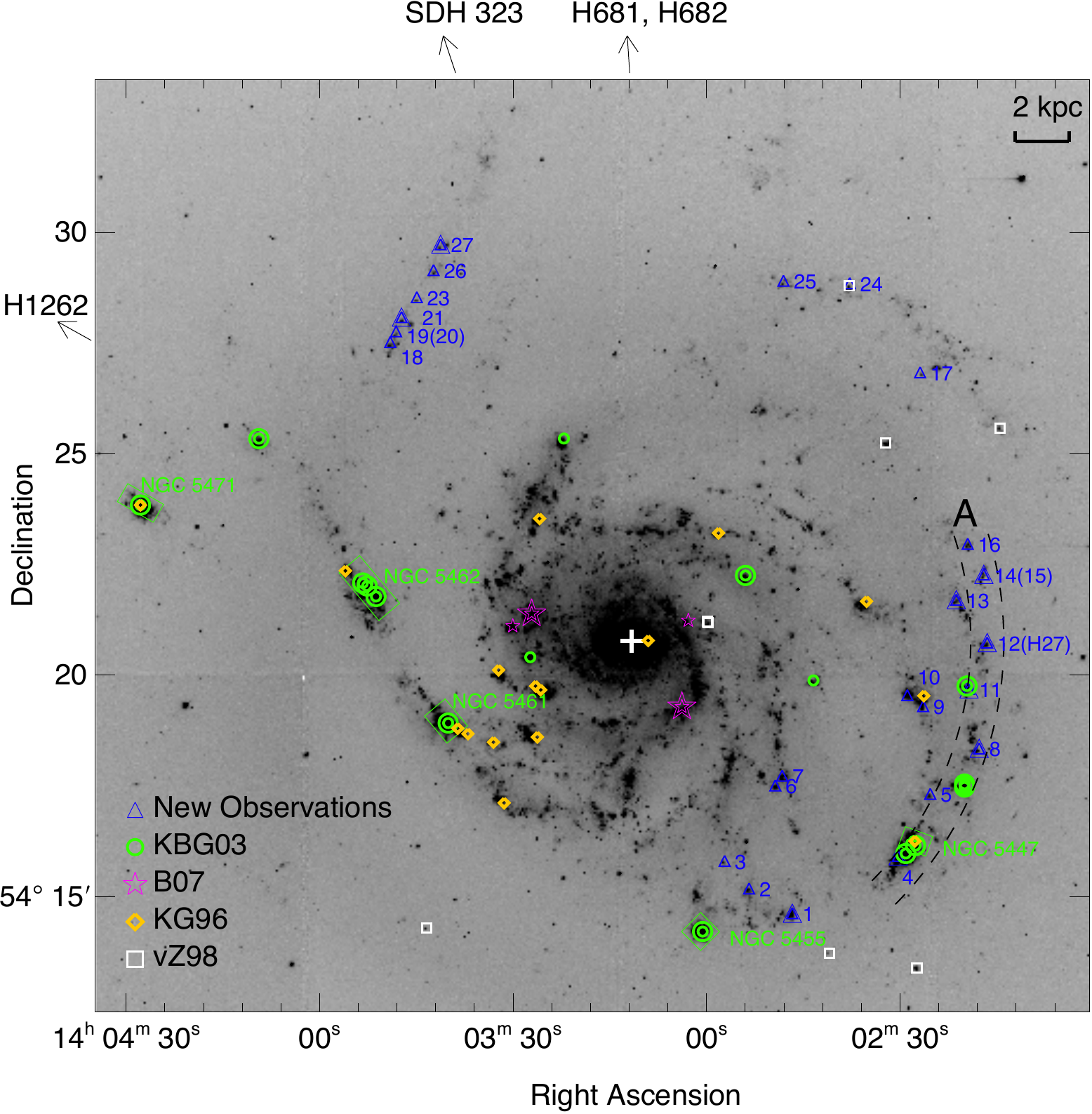} 
    \caption{Location of the \hii\ regions in our study on a narrow-band H$\alpha$ image of M101 (from the Canada-France-Hawaii Telescope data archive). The white cross marks the  center of the galaxy.
        The symbols indicate the different data sources: blue triangles - new observations; green circles - \citet{Kennicutt:2003};  magenta stars -  \citet{Bresolin:2007}; orange diamonds -  \citet{Kennicutt:1996}; 
 white squares -  \citet{van-Zee:1998a}. \hii\ regions with \Oiii$\lambda4363$ detections are labeled by doubled line symbols. Note that four distant objects (H1262, H681, H682 and SDH323) are outside the field of view (the arrows indicate the directions to these nebulae). The two curved dashed lines in the west mark the position of  `arc A'.\label{fig:f1}}
     
\end{figure*}

We obtained multi-object spectroscopy of M101 during the nights of May 7-8, 2010 with the \mbox{R-C}  spectrograph
at the Mayall 4-m telescope of the Kitt Peak National Observatory, using multi-object masks in five different fields, with 2 arcsec-wide slits.   The data were acquired at airmasses smaller than 1.25, to minimize the impact of differential atmospheric dispersion. In each field we integrated for $3\times2400$\,s  using the KPC-10A grating (2.75~{\AA} pixel$^{-1}$), which yielded a $\sim$7.5~{\AA} FWHM spectral resolution and  covering approximately the wavelength range between 3600~{\AA} and 7500~{\AA}. 

Standard IRAF\footnote{IRAF is distributed by the National Optical Astronomy Observatory, which is operated by the Association of Universities for Research in Astronomy, Inc., under cooperative agreement with the National Science Foundation.} tasks, in combination with the PyRAF\footnote{PyRAF is a product of the Space Telescope Science Institute, which is operated by AURA for NASA} command language, were used for bias subtraction, flat-field correction,  cosmic ray removal, spectral extraction, image co-addition and wavelength calibration (using He+Ne+Ar lamp frames). We extracted one-dimensional spectra for each target \hii\ region after executing all the standard reduction procedures. Three standard stars (Feige~34, Feige~67 and BD+$28\,4211$) were observed several times each night for the flux calibration. 

Our  observations yielded spectra for 28  \hii\ regions.
Their celestial coordinates, galactocentric distances normalized to the isophotal radius $R_{0}$,  position angles $\phi$ relative to the galactic center and identifications from the  \citet{Hodge:1990} catalog, are presented in Table~\ref{table:positions} (where objects are listed in order of increasing declination).
We have adopted the following parameters:  center coordinates RA = 14:03:12.5, Dec = +54:20:56 (J2000),  position angle of the major axis = 37 deg,  inclination angle = 18 deg \citep{de-Blok:2008}, distance = 6.85 Mpc \citep{Freedman:2001} and  isophotal radius $R_0$ $=$ 14\farcm4 \citep{de-Vaucouleurs:1991}, corresponding to 28.7~kpc. 
Fig.~\ref{fig:f1} shows  the location of the target \hii\ regions in M101. \hii\ regions for which  we detected the \Oiii$\lambda$4363 auroral line are labeled by double symbols, and we identify the different data sources  considered in this paper with different symbols. 


\subsection{Line flux measurement} \label{s:flux}


\begin{deluxetable*}{cccccc}
\tablecolumns{6}
\small
\tablewidth{0pt}
\tabletypesize{\scriptsize}
\tablecaption{Observed \hii\  region sample\label{table:positions}}
\tablehead{
\colhead{ID\tablenotemark{a}} & 
\colhead{R.A.\tablenotemark{b}} &
\colhead{Decl.\tablenotemark{c}} & 
\colhead{$R/R_{0}$\tablenotemark{d}} &
\colhead{$\phi$\tablenotemark{e}} & 
\colhead{Other ID\tablenotemark{f} }
 \\
 \colhead{ }&
 \colhead{(J2000.0)} & 
 \colhead{(J2000.0)} & 
 \colhead{ } &
 \colhead{(degree)} &
 \colhead{} 
}
\startdata 
1&   14 02 46.92  &   54 14 50.1  &  0.50 & 212  &  H219 \\    	           
2&   14 02 53.68  &   54 15 22.8  &  0.43 & 206  &  H321 \\			  
3&   14 02 57.54  &   54 15 59.8  &  0.37 & 204  &  H370 \\                         
4&   14 02 30.61  &   54 16 09.7  &  0.54 & 232  &  H149 \\			   
5&   14 02 25.73  &   54 17 34.2  &  0.53 & 244  &  H103 \\			   
6&   14 02 49.83  &   54 17 43.0  & 0.32 & 226  &  H260 \\			   
7&   14 02 48.78  &   54 17 56.3  & 0.32 & 229  &  H237 \\			   
8&   14 02 18.22  &   54 18 36.6  &  0.58 & 254  &  H41  \\    			   
9&   14 02 27.01  &   54 19 33.0  &  0.48 & 258  &  H120 \\ 			   
10&   14 02 29.49  &   54 19 47.7  &   0.45 & 260  &  H140 \\			   
11&   14 02 19.92  &   54 19 57.1  &   0.55 & 263  &  H59  \\ 			   
12&   14 02 16.85  &   54 21 00.1  &   0.58 & 271  &  H27  \\			   
13&   14 02 22.06  &   54 21 57.7  &   0.54 & 278  &  H79  \\			   
14&   14 02 17.83  &   54 22 32.4  &   0.59 & 282  &  H37\tablenotemark{g}\\  	     	   
15&   14 02 17.83  &   54 22 32.4  &   0.59 & 282  &  H37\tablenotemark{g}\\		   
16&   14 02 20.39  &   54 23 13.7  &   0.58 & 287  &  H68  \\			   
17&   14 02 27.50  &   54 27 08.0  &   0.66 & 313  &  H125 \\			   
18&   14 03 50.86  &   54 27 35.9  &   0.61 &  40  &  H1151\\			   
19&   14 03 49.98  &   54 27 50.5  &   0.61 &  38  &  H1148\tablenotemark{g}\\	   
20&   14 03 49.98  &   54 27 50.5  &   0.61 &  38  &  H1148\tablenotemark{g}\\	  
21&   14 03 49.18  &   54 28 10.0  &   0.63 &  36  &  H1146\\			
22&   14 04 49.80  &   54 28 14.0  &   1.12 &  63  &  H1262\\			
23&   14 03 46.84  &   54 28 37.3  &   0.64 &  33  &  H1137\\			
24&   14 02 39.30  &   54 29 04.9  &   0.69 & 329  &  H188 \\			
25&   14 02 49.65  &   54 29 07.0  &   0.64 & 338  &  H253 \\			
26&   14 03 44.30  &   54 29 14.4  &   0.66 &  29  &  H1121\\			
27&   14 03 43.26  &   54 29 49.4  &   0.69 &  27  &  H1118\\			
28&   14 03 13.70  &   54 35 47.0  &   1.05 &   1  &  H681			   	
\enddata
\tablenotetext{a}{\hii\ region identification (in order of increasing declination)}
\tablenotetext{b}{Right ascension. Units are hours, minutes and seconds} 
\tablenotetext{c}{Declination. Units are degrees, arcminutes and arcseconds}
\tablenotetext{d}{Deprojected galactrocentric distance in units of the isophotal radius, $R_0$ = $14\farcm4$,  taken from \citet{de-Vaucouleurs:1991} }
\tablenotetext{e}{Position angle relative to the galaxy center, in degrees}
\tablenotetext{f}{Identification from \citet{Hodge:1990}}  
\tablenotetext{g}{Two objects in the same slit}

\end{deluxetable*}

\begin{deluxetable*}{ccccccccccc}
\centering
   \tablecaption{Dereddened line fluxes, temperatures and oxygen abundances from the direct method 
   \label{table:linefluxes}}
\tabletypesize{\scriptsize}  

  \tablehead{
\colhead{ID} &
\colhead{Name} &
\colhead{\Oii} &
\colhead{\Oiii}  &
\colhead{[O III]}  &
\colhead{\Nii}& 
\colhead{\Sii} & 
\colhead{F(H$\beta$)} & 
\colhead{c(H$\beta$)} & 
\colhead{T\Oiii} &
\colhead{12+log(O/H)} 
\\
\colhead{}&
\colhead{}& 
\colhead{\lin3727} &
\colhead{\lin4363} &
\colhead{\lin5007} &
\colhead{\lin6583} &
\colhead{\llin6717,6731} &
\colhead{(erg $s^{-1}$ $cm^{-2}$)} &
\colhead{(mag)} &
\colhead{(K)} &
\colhead{}
}
\startdata    
1& H219 &  413  $\pm$  21 &  0.91  $\pm$  0.05 & 178  $\pm$   8  & 36.0  $\pm$  1.5  &  41.3 $\pm$  1.3 &   3.8 $\times10^{-14}$   & 0.31  & 9420  $\pm$   258  &   8.42 $\pm$   0.08  \\
2& H321 &  264  $\pm$  13 &      \ldots        & 237  $\pm$  11  & 29.5  $\pm$  1.2  &  24.7 $\pm$  0.7       &   3.3 $\times10^{-14}$   & 0.43  &     \ldots         &     \ldots  \\         
3& H370 &  283  $\pm$  14 &      \ldots        & 129  $\pm$   6  & 78.3  $\pm$  2.8  &  54.4 $\pm$  1.4        &   1.6 $\times10^{-14}$   & 0.53  &     \ldots         &     \ldots  \\         
4& H149 &  238  $\pm$  12 &  1.68  $\pm$  0.08 & 329  $\pm$  15  & 35.6  $\pm$  1.5  &  39.4 $\pm$  1.2 &   1.6 $\times10^{-13}$   & 0.35  & 9411  $\pm$   247  &   8.40 $\pm$   0.08  \\  
5& H103 &  343  $\pm$  17 &      \ldots        & 158  $\pm$   7  & 42.6  $\pm$  2.0  &  38.6 $\pm$  1.3        &   1.3 $\times10^{-14}$  & 0.18  &     \ldots         &     \ldots      \\       
6& H260 &  245  $\pm$  12 &      \ldots        & 207  $\pm$   9  & 36.6  $\pm$  1.4  &  35.4 $\pm$  1.0        &   3.6 $\times10^{-14}$   & 0.45  &     \ldots         &     \ldots    \\       
7& H237 &  193  $\pm$  10 &      \ldots        & 106  $\pm$   5  & 47.5  $\pm$  1.7  &  38.5 $\pm$  1.0        &   8.2 $\times10^{-14}$   & 0.51  &     \ldots         &     \ldots    \\       
8& H41   &  169   $\pm$   9 &  3.92  $\pm$  0.18 & 476  $\pm$  21  & 8.4  $\pm$  0.4  &  15.8 $\pm$  0.5    &   3.7 $\times10^{-14}$ & 0.28   & 10894  $\pm$   327 &   8.25 $\pm$   0.08  \\  
9& H120 &  419  $\pm$  21 &      \ldots        &  82  $\pm$   4  & 63.3  $\pm$  2.8  &  86.9 $\pm$  2.8         &   1.9 $\times10^{-14}$ & 0.26  &     \ldots         &     \ldots           \\  
10&H140&  352  $\pm$  18 &      \ldots        & 145  $\pm$   7  & 57.0  $\pm$  2.5  &  67.1 $\pm$  2.1        &   3.1 $\times10^{-14}$ & 0.29  &     \ldots         &     \ldots       \\      
11&H59  &  305   $\pm$  15 &  2.30  $\pm$  0.11 & 290  $\pm$  13  & 17.2  $\pm$  0.8  &  25.9 $\pm$  0.9  &   2.1 $\times10^{-14}$ & 0.16  &10769  $\pm$   322  &   8.24 $\pm$   0.08  \\  
12&H27  &  341   $\pm$  17 &  3.64  $\pm$  0.17 & 251  $\pm$  11  & 21.6  $\pm$  1.0  &  58.0 $\pm$  2.0   &   2.1 $\times10^{-14}$ & 0.16  &13401  $\pm$   499  &   7.96 $\pm$   0.08  \\  
13&H79  &  226   $\pm$  11 &  3.04  $\pm$  0.15 & 330  $\pm$  15  & 15.5  $\pm$  0.7  &  27.1 $\pm$  0.9   &   1.7 $\times10^{-14}$ & 0.25  &11313  $\pm$   563  &   8.14 $\pm$   0.11  \\  
14&H37  &  236   $\pm$  12 &      \ldots        & 333  $\pm$  15  & 19.2  $\pm$  0.9  &  22.1 $\pm$  0.8         &   8.6 $\times10^{-15}$  & 0.19  &     \ldots         &     \ldots           \\  
15&H37  &  234   $\pm$  12 &  3.31  $\pm$  0.16 & 387  $\pm$  17  & 14.9  $\pm$  0.6  &  28.5 $\pm$  0.8   &   2.5 $\times10^{-14}$ & 0.38  &11043  $\pm$   344  &   8.22 $\pm$   0.08  \\  
16&H68  &  391   $\pm$  19 &      \ldots        & 171  $\pm$   7  & 33.8  $\pm$  1.5  &  75.6 $\pm$  2.5          &   1.1 $\times10^{-14}$ & 0.26  &     \ldots         &     \ldots         \\    
17&H125&  350  $\pm$  18 &      \ldots        & 327  $\pm$  15  & 29.0  $\pm$  1.4  &  33.3 $\pm$  1.3         &   1.3 $\times10^{-15}$ & 0.15  &     \ldots         &     \ldots           \\  
18&H1151 &  397  $\pm$  21 &      \ldots       & 158  $\pm$   7  & 25.0  $\pm$  1.0  &  35.4 $\pm$  1.1          &   2.1 $\times10^{-15}$ & 0.43  &     \ldots         &     \ldots           \\  
19&H1148 &  177  $\pm$   9 &      \ldots       & 293  $\pm$  13  & 17.5  $\pm$  0.8  &  25.6 $\pm$  0.8          &   7.9 $\times10^{-15}$ & 0.28  &     \ldots         &     \ldots           \\  
20&H1148 &  199  $\pm$  10 &      \ldots       & 383  $\pm$  17  & 11.0  $\pm$  0.4  &  14.5 $\pm$  0.4         &   1.5 $\times10^{-14}$ & 0.40  &     \ldots         &     \ldots           \\  
21&H1146 &  212  $\pm$  11 &  3.07  $\pm$ 0.15 & 457  $\pm$  20  & 17.7  $\pm$  0.8  &  27.1 $\pm$  0.9   &   9.6 $\times10^{-15}$ & 0.27  &10219  $\pm$   299  &   8.35 $\pm$   0.08  \\  
22&H1262 &  200  $\pm$  10 &      \ldots       & 316  $\pm$  14  & 13.0  $\pm$  0.7  &  31.6 $\pm$  1.1          &   3.4 $\times10^{-15}$ & 0.22  &     \ldots         &     \ldots           \\  
23&H1137 &  378  $\pm$  19 &      \ldots       &  25  $\pm$   1  & 43.7  $\pm$  2.0  &  87.5 $\pm$  2.9           &   2.3 $\times10^{-15}$ & 0.25  &     \ldots         &     \ldots           \\  
24&H188&  682  $\pm$  34 &      \ldots        & 164  $\pm$   7  & 23.8  $\pm$  1.1  &  58.7 $\pm$  2.0          &   3.9 $\times10^{-15}$ & 0.23  &     \ldots         &     \ldots          \\   
25&H253&  430  $\pm$  22 &      \ldots        &  73  $\pm$   3  & 40.8  $\pm$  1.9  &  84.6 $\pm$  2.9           &   3.6 $\times10^{-15}$ & 0.17  &     \ldots         &     \ldots          \\   
26&H1121 &  336  $\pm$  17 &      \ldots       & 217  $\pm$  10  & 29.3  $\pm$  1.3  &  47.7 $\pm$  1.6         &   5.7 $\times10^{-15}$ & 0.26  &     \ldots         &     \ldots           \\  
27&H1118 &  222  $\pm$  11 &  5.97  $\pm$ 0.29 & 559  $\pm$  25  & 8.9  $\pm$  0.4  &  11.4 $\pm$  0.4    &   6.4 $\times10^{-15}$ & 0.27   &11917  $\pm$   402  &   8.21 $\pm$   0.08  \\  
28&H681 &  232  $\pm$  12 &  4.36  $\pm$  0.20 & 260  $\pm$  12  & 9.7  $\pm$  0.3  &  27.7 $\pm$  0.7     &   2.6 $\times10^{-14}$  & 0.59  &14237  $\pm$   559   &   7.80 $\pm$   0.09  

\enddata
\tablecomments{  
Line fluxes are normalized to H$\beta$$=$100, after correcting for reddening. F(H$\beta$) is the measured H$\beta$ flux, corrected for extinction.
}

\end{deluxetable*}

The emission line intensities were measured with the {\it splot} routine in IRAF, by integrating the flux under the line profiles between two continuum points selected interactively. A multi-gaussian profile fit was performed if evident line blending occurred.

A correction for interstellar reddening was performed adopting the \citet{Howarth:1983} analytical formulation of the \citet{Seaton:1979} law, assuming a total-to-selective extinction ratio R$_V$ $=$ A$_V$/E$_{B-V}$ = 3.1. By comparing the measured intensities of H$\alpha$ and H$\gamma$, relative to H$\beta$, to the case B theoretical values taken from \citet{Storey:1995} calculated at the electron temperatures determined from the auroral lines (or 10,000~K if these lines were not detected), we obtained the reddening coefficient c(H$\beta$). The Balmer lines ratios (H$\alpha$/H$\beta$, H$\gamma$/H$\beta$) were corrected for  underlying stellar absorption in an iterative manner. 

The resulting reddening-corrected line flux measurements,  normalized to H$\beta$ = 100, are tabulated in Table~\ref{table:linefluxes}. The errors in  these fluxes were estimated from  the uncertainties in the line intensity measurements and the flux calibration, the  scatter of the continuum near the emission lines, and the uncertainties in the extinction coefficient. We note that, in general, the \hii\ regions in our sample are extended, and are not fully covered by the width of our slits. Thus, the H$\beta$ fluxes reported in Table~2 should be considered lower limits to the real fluxes.  

\subsection{Electron temperatures and oxygen abundances}\label{s:TandZ}
Electron temperatures  and  oxygen abundances were obtained adopting a
two excitation zone structure for the \hii\ regions, characterized by the temperatures T\Oii\ and T\Oiii.

To calculate the total oxygen abundances, we made the usual assumption: 
\begin{equation}
\rm O/H=(O^{+}+O^{++})/H^{+}
\end{equation}

The electron temperature ($T_e$) for the O$^{++}$-emitting region (T\Oiii) and for the O$^{+}$-emitting region (T\Oii) were derived from the auroral-to-nebular line intensity ratios (\Oiii$\lambda$4363/$\lambda\lambda$4959,
5007 and \Oii$\lambda$7325/$\lambda3727$), when available, with the {\it nebular} module \citep{Shaw:1995} of the STSDAS package running under IRAF. For nebulae without auroral line detections we adopted $T_e = 10^4$~K.
One or both of the \Oiii$\lambda4363$ and \Oii$\lambda$7325 auroral lines could be measured in 10 out of the 28 \hii\ regions in our sample: seven objects with both lines detected, three with only \Oiii$\lambda4363$ detected and one with only \Oii$\lambda7325$ detected. We 
compared our measured T\Oii\ values with those obtained from the theoretical relation obtained by
\citet{Garnett:1992}:

\begin{equation}\label{garnetteq}
{\rm T[O~{\scriptstyle II}] = 0.70~T[O~{\scriptstyle III}] + 3000~K}
\end{equation}

\noindent
and found that the two were consistent with each other, within the uncertainties. However, since the T\Oii\ values obtained from the auroral-to-nebular line ratios have significantly larger errors (between 300~K and 900~K) than those propagated from T\Oiii\ through Eq.~\ref{garnetteq} (170~K to 400~K), a result already noticed by \citet{Kennicutt:2003}, we decided to adopt the T\Oii\ values calculated via Eq.~\ref{garnetteq}. In the end, this choice has a limited impact on the abundance determination, since
the O/H ratios obtained by adopting one or the other set of T\Oii\ values  differ only by a few hundreds of a dex (0.07 dex at most), except for H59, where the difference is 0.13 dex. 

The measured $T_e$ values of the O$^{++}$-emitting region are presented in Table~\ref{table:linefluxes}. We assume, as indicated by the \sii\lin6717/\lin6731 line ratios, that all \hii\ regions are in the low density limit ($n_{e}$~$<$~10$^2$ cm$^{-3}$). 

\smallskip
We derived the oxygen abundances both from  the measurement of the electron temperature (`direct' method or $T_e$ method), obtained as explained above, and strong-line diagnostics. %
Strong-line diagnostics were put forward to obtain the abundances in \hii\ regions when the temperature-sensitive lines are not accessible (e.g.~in metal-rich regions or high redshift galaxies), by using combinations of more easily measured strong emission lines as indicators of the abundance, such as the popular \Oii$\lambda$3727$+$\Oiii$\lambda\lambda$4959, 5007)/H$\beta$ line ratio, otherwise known as $R_{23}$ \citep{Pagel:1979}. As is well known, strong-line indicators, like $R_{23}$, can suffer from degeneracy when calculating the chemical abundances, and abundances derived from different strong-line methods and/or calibrations can vary systematically  up to $\Delta$log(O/H)\,=\,0.7 dex \citep{Kewley:2008}.

\subsection{The enlarged sample}\label{s:sample}

 \begin{figure*}
\centering
    \includegraphics[width=1.\columnwidth]{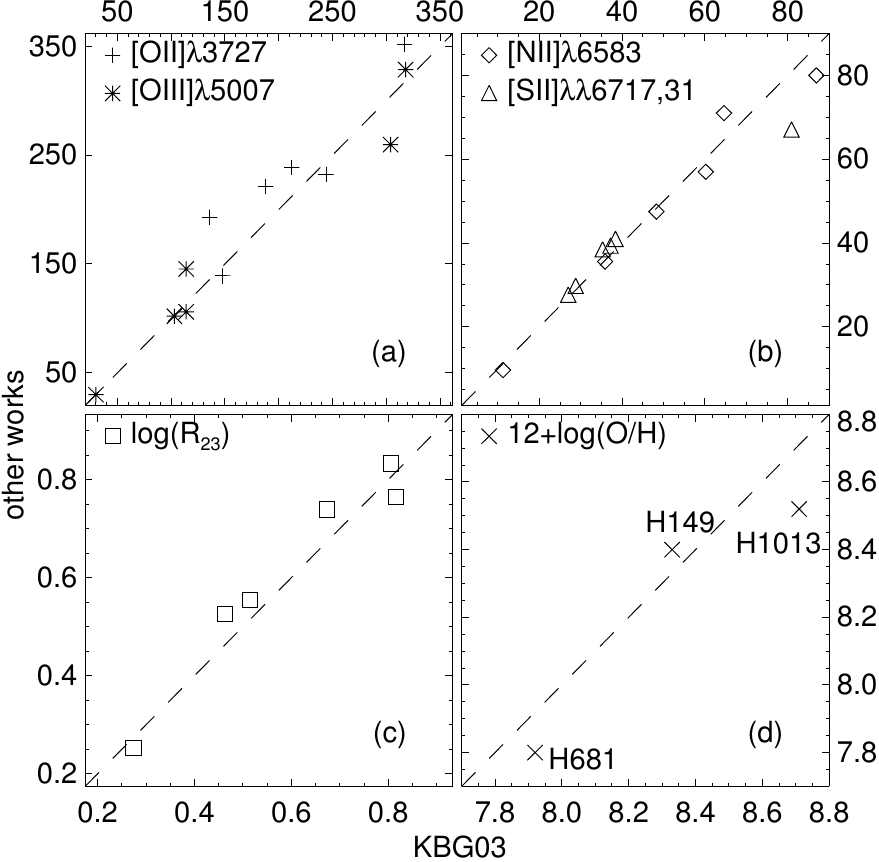}
    \caption{Comparison of line fluxes and derived quantities for \hii\ regions  from our new observations and those by \citet[KBG03]{Kennicutt:2003}. {\it (a)}  \Oii$\lambda3727$ (crosses) and \Oiii$\lambda$5007 (asterisks); {\it (b)}  \Nii$\lambda6583$ (diamonds) and  \Sii$\lambda\lambda$6717,6731 (triangles); {\it (c)} log(R$_{23}$), {\it (d)} T$_e$-based oxygen abundances. The dashed line represents the one-to-one correlation.
      \label{fig:f2} }
      \end{figure*}
      
   
\begin{deluxetable*}{llllllll}
\tablecaption{Comparison with previous observations\label{table:common}}
\tablehead{
\colhead{} &
\colhead{}&
\colhead{H1013\tablenotemark{a}}&
\colhead{H972\tablenotemark{a}}&
\colhead{H237\tablenotemark{b}}&
\colhead{H149\tablenotemark{b}}&
\colhead{H140\tablenotemark{b}}&
\colhead{H681\tablenotemark{b}}
}
\startdata
                                                       &     KBG03  &      188         &              148	    &  	136		&  	212		&  	317		&  	244  \\
\Oii$\lambda$3727		                                    &    B07   &      221         &           139	& 	\ldots	& 	\ldots	& 	\ldots	& 	\ldots\\
            				             &    New  &     \ldots       &            \ldots& 	192.8	& 	238.5	& 	351.8	& 	232\\
\\                                                        
\Oiii$\lambda$5007           				         &    KBG03  &     103          &            30	& 	114		& 	318		& 	114		& 	304\\
                                                       &    B07   &     102          &           30	& 	\ldots	& 	\ldots	& 	\ldots	& 	\ldots\\
                                                        &    New  &     \ldots       &          \ldots	& 	106.1	& 	328.6	& 	145.4	& 	259\\
\\
\Nii$\lambda$6583                  				&     KBG03 &      64.6        &           86.9	& 	48.3		& 	35.9		& 	60.2		& 	11.3\\
                                                           &     B07  &      71           &          80	& 	\ldots	& 	\ldots	& 	\ldots	& 	\ldots\\
                                                        &     New &      \ldots      &         \ldots	& 	47.5		& 	35.6		& 	57		& 	9.7\\
\\
\Sii$\lambda\lambda$6717,6731						  &    KBG03  &    28.8          &      38.4	 &	35.3		 &	37.2		 &	80.9		 &	27\\
                                                         &    B07   &    29.8          &         41	 &	\ldots	 &	\ldots	 &	\ldots	 &	\ldots\\
                                                         &    New  &     \ldots       &       \ldots	 &	38.5		 &	39.4		 &	67.1		 &	27.7\\
\\
\Oiii$\lambda$4363                 				     &    KBG03  &      \ldots       &          \ldots  &   \ldots		 &       1.8		&   	\ldots	  & 	\ldots\\
                                                        &    B07   &    0.24          &        \ldots	 &	\ldots	 &	\ldots	 &	\ldots	 &	\ldots\\
                                                         &   New   &    \ldots	 &	   \ldots	 &	\ldots	 &	1.7		 &	\ldots	 &	\ldots\\
\\
R$_{23}$                                          &   KBG03   &    3.27          &           1.88	 &	2.90		 &	6.40		 &	4.71		 &	6.54\\
                                                         &   B07    &    3.58          &            1.79	 &	\ldots	 &	\ldots	 &	\ldots	 &	\ldots\\
                                                         &   New   &    \ldots        &           \ldots &	3.36		 &	6.81		 &	5.48		 &	5.82\\
\\
12+log(O/H)$_{T_{e}}$     				  &   KBG03   &     8.71         &             \ldots   &     \ldots	 &	8.33		 &	\ldots	 &	7.92\\
                 					    &   B07    &    8.52          &              \ldots &      \ldots	 &	\ldots	 &	\ldots	 &	\ldots\\
               				             &   New   &    \ldots        &              \ldots  &       \ldots	 &	8.40		 &	\ldots	 &	7.80
 \enddata
 
\tablecomments{Six \hii\ regions are in common between KBG03, B07 and our new observations. }
 \tablenotetext{a}{For H1013 and H972, the B07 line fluxes were adopted to construct the enlarged \hii\ region  sample (see text).}
 \tablenotetext{b}{For H237, H149, H140 and H681 the newly-determined line fluxes from our observations were used for the enlarged sample.}
 \end{deluxetable*}

In order to increase the sample size and to cover a larger radial and azimuthal range, we also collected data from additional sources:

\vspace{2.5 mm}
$\bullet$ \citet[][=\,KG96]{Kennicutt:1996} presented spectroscopic data for 41 \hii\ regions, including some distinct knots in large complexes (NGC~5462, NGC~5447 and NGC~5471), which could be used as control samples to examine the intrinsic metallicity variation, if we assume that within their volume the oxygen abundance is homogeneous.

\vspace{2.5 mm}
$\bullet$ \citet[=\,KBG03]{Kennicutt:2003} observed 26 \hii\ regions, with 23 targets in common with KG96; 17 of those 23 observations yielded \Oiii$\lambda4363$ measurements. In addition, they also observed 3 new objects:  H70, H71 and SDH323 (following the designation in KBG03), all of them with \Oiii$\lambda$4363 measurements.

\vspace{2.5 mm}
$\bullet$ \citet[][=\,B07]{Bresolin:2007} observed 4 \hii\ regions (H1013, H493, H507 and H972) in the inner (central 3\arcmin), metal-rich zone of M101; two (H1013 and H493) provided reliable $T_e$ measurements. 

\vspace{2.5 mm}
$\bullet$ \citet[=\,vZ98] {van-Zee:1998a} published spectroscopic data for 13 \hii\ regions  in M101. These authors did not  follow the object identification from the catalog of \citet{Hodge:1990}. Based on their map of slit positions, we identified the corresponding  objects in \citet{Hodge:1990}.
\vspace{2.5 mm}

In assembling our catalog of line fluxes for \hii\ regions in M101, we adopted the more accurate line fluxes given by KBG03 and B07 for the 26 objects in common with KG96. Similarly, for H67 and H188, which were observed by vZ98, we adopted our new data or those by KBG03.

Since in KG96 and vZ98 only the sum of the \Oiii $\lambda\lambda$4959,5007 and \Nii $\lambda\lambda$6548,6583 line fluxes are given, we calculated the individual line fluxes of the lines in the doublets using the theoretical ratios \citep{Storey:2000}.

Table~\ref{table:common} presents line fluxes from different authors (KBG03 and B07) for the six objects in common with our new dataset of 28 \hii\ regions (labeled as ``New" in the table). In Fig.~\ref{fig:f2} we
display  a comparison of line fluxes and derived quantities ($R_{23}$, O/H).

The agreement between the different studies is generally satisfactory.
There is only one object, H140, with a considerable difference ($\sim$20\%) in the \sii\ flux between our new observations and those by KBG03. Fig.~\ref{fig:f2}(c) indicates that $R_{23}$ values are consistent between the different studies, with differences $<0.12$ dex. There are only three objects in common that also have an \Oiii$\lambda4363$-based oxygen abundances. 
For H1013, B07 provided an oxygen abundance 0.21 dex lower than KBG03. We adopted the value from B07, obtained from deeper data, and which corresponds to an O$^{++}$ zone temperature directly derived from \Oiii $\lambda4363/(\lambda4959 + \lambda 5007)$, while in KBG03 it was derived from T\Siii. Finally, we adopted the H972 measurements  from B07, while for the remaining four \hii\ regions included in Table~\ref{table:common}
our new observations were adopted.

Our final sample comprises 79 \hii\ regions,  with 28 objects from our new observations, 20 from KBG03, 4 from B07, 16 from KG96 and 11 from vZ98. In this merged sample, there are 29 measurements of the \Oiii$\lambda4363$ auroral line  (10 from our new observations, 17 from KBG03 and 2 from B07). 
In order to ensure consistency in the analysis, we recalculated the $T_e$ and 
the direct oxygen abundances when the \Oiii$\lambda$4363 line was measured.

\begin{figure}
\centering
    \includegraphics{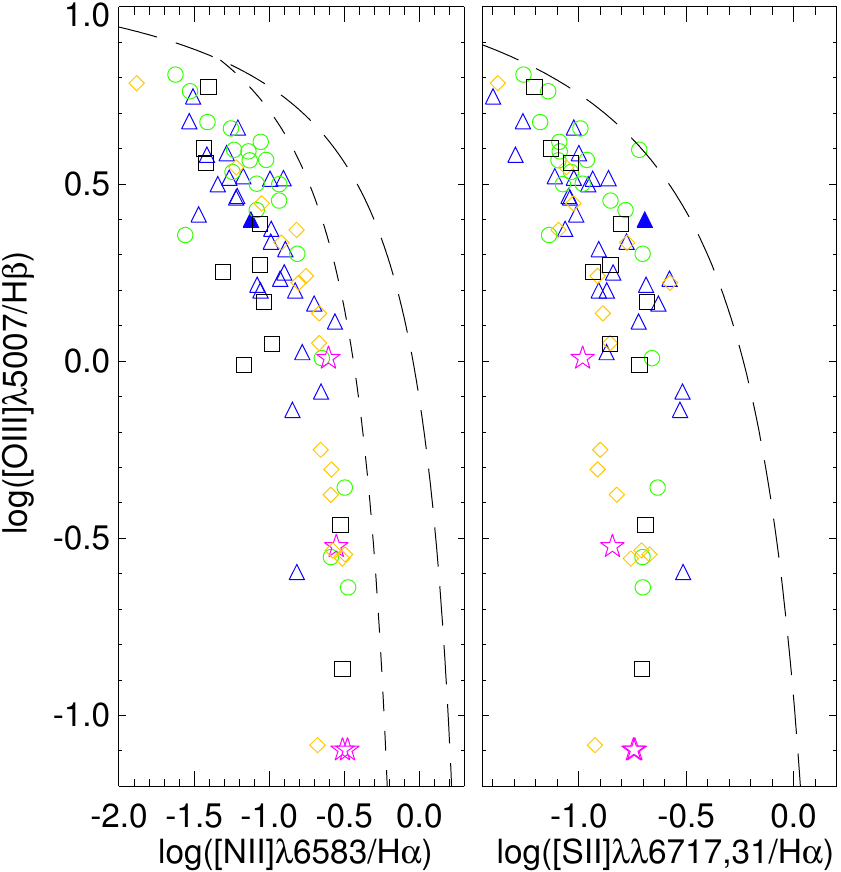}
    \caption{Nebular diagnostic diagrams showing the excitation sequence for  our \hii\ region sample:
    log(\Nii $\lambda$6583/H$\alpha$) {\it (left)} and log(\Sii $\lambda\lambda$6717,6731/H$\alpha$) {\it (right)} as a function of
log(\Oiii $\lambda$5007/H$\beta$).  The curves represent   upper boundaries for star-forming regions ionized by hot stars from \citet[long-dashed line]{Kewley:2001} and \citet[short-dashed line]{Kauffmann:2003}. The filled blue triangles mark the positions of H27 in our new observations. Symbols as in Fig.~\ref{fig:f1}. \label{fig:f3}}
      
      \end{figure}

In Fig.~\ref{fig:f3} we examine the excitation properties of our \hii\ region sample by plotting the classic \Oiii $\lambda$5007/H$\beta$ \vs\ \Nii $\lambda6583$/H$\alpha$ and \Oiii$\lambda5007$/H$\beta$ \vs\ \Sii $\lambda$6717,6731/H$\alpha$ diagnostic diagrams (\citealt{Baldwin:1981}\,=\,BPT).

We include in the plot the  boundaries between different photoionization sources (star-forming regions \vs\ AGN) by \citet[shown by the long-dashed line]{Kewley:2001}
and \citep[ short-dashed line]{Kauffmann:2003}.
\citet{Kewley:2001} used stellar population synthesis  and photoionization models to create the `maximum starburst line' on the BPT diagrams. A modification to this classification was provided by \citet{Kauffmann:2003} to include an empirical boundary line between pure star-forming galaxies and Seyfert-\hii\ composite objects. 
As shown in Fig.~\ref{fig:f3}, according to these criteria all the objects in our sample are located in the pure star-forming region of the diagram. 

 


 \begin{figure*}
\centering
    \includegraphics{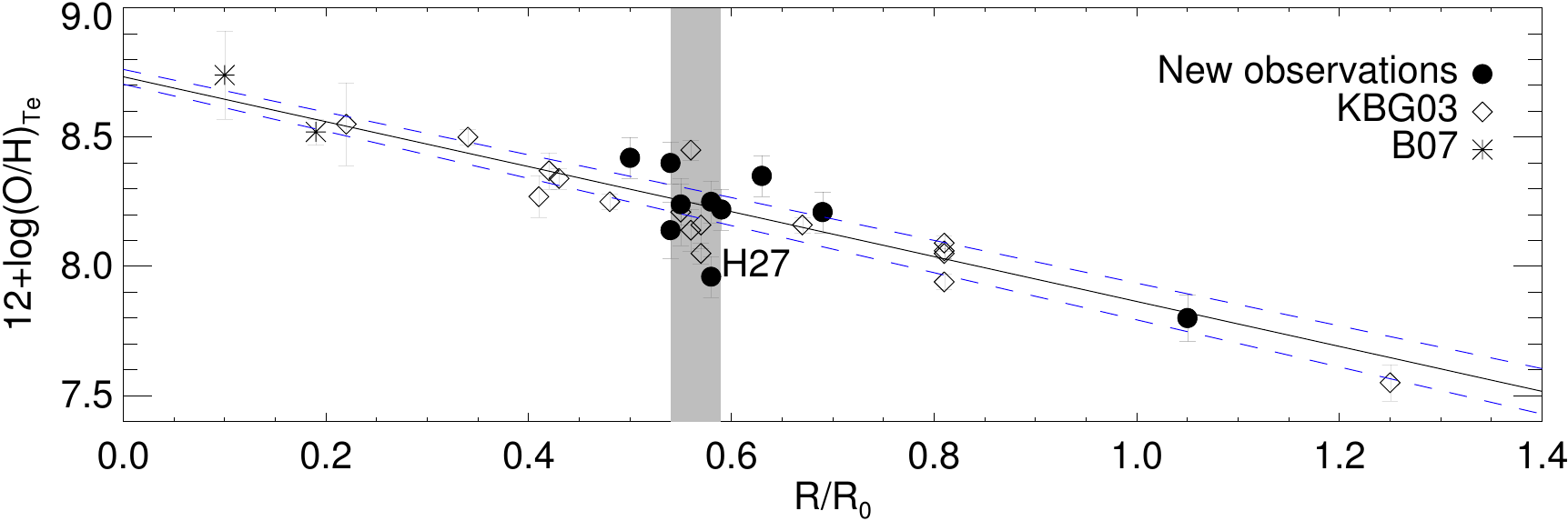}
    \caption{Radial oxygen abundance gradient determined from direct electron temperature measurements. The filled circles represent our new observations. The \hii\ regions from KBG03 and B07 are labeled by the diamond and the asterisk symbols, respectively. 
    The solid line is the weighted linear least-squares fit of all 29 \hii\ regions in the figure, with the 1$\sigma$ uncertainty represented by the two blue dashed lines. 
    The grey shadowed region is the portion where we have the most data points in a radial range of $\sim$1.4 kpc ($R/R_{0}$\,=\,0.54 to 0.59). 
        H27 is the \hii\ region with the most significant oxygen abundance discrepancy relative to the linear fit.
      \label{fig:f4} }
      \end{figure*}

\begin{figure}
\centering
    \includegraphics{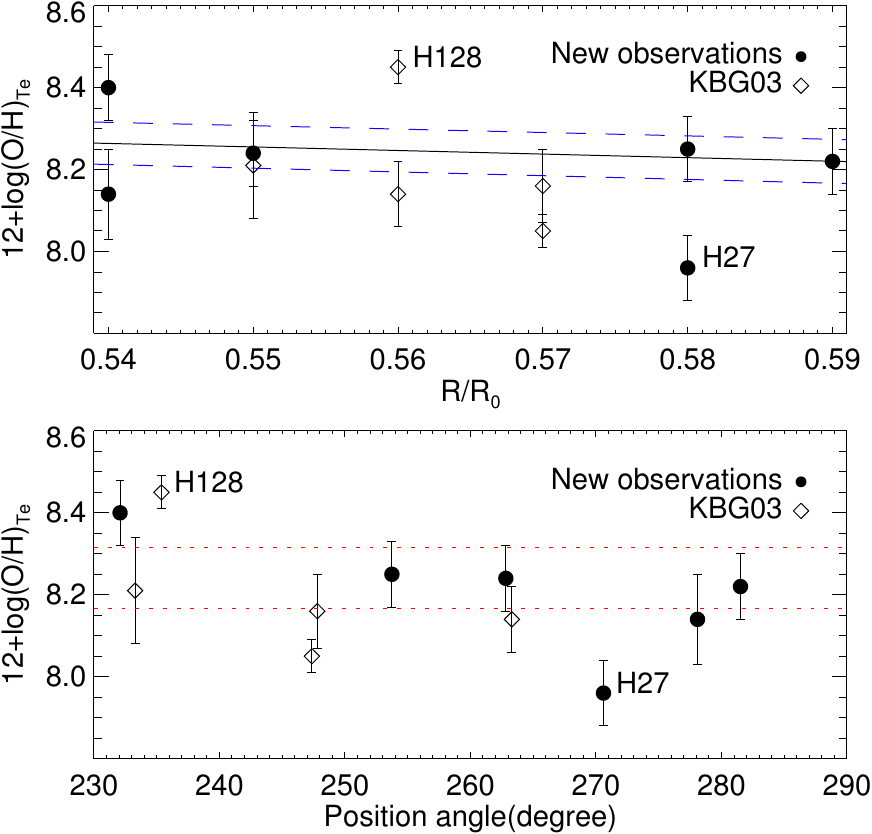}
    \caption{Enlargement of the gray shadowed region in Fig.~\ref{fig:f4}. 
    The upper panel shows the relation between the oxygen abundance and the radial distance range $R/R_{0}$\,=\,0.54 to 0.59. The blue dashed line has the same meaning as in Fig.~\ref{fig:f4}. 
    The bottom panel is the variation of the oxygen abundance with  position angle. The region between the two red dotted lines is the maximum variation of the oxygen abundance due to the change of  radial distance, calculated from Eq.~\ref{eq:radialte}.
      \label{fig:f5} }
      \end{figure}

\section{Radial abundance gradient and local azimuthal variations}\label{s:radial}
The first  measurements of a radial abundance gradient from  \hii\ regions in M101 can be traced back to more than 40 years ago \citep{Searle:1971,Smith:1975}. 
Based on our new spectroscopic observations, in combination with the  work published by other authors, as presented in
Sect.~\ref{s:observations}, we are in the position to better assess the spatial distribution of the oxygen abundance in the disk of this galaxy. Given the relatively dense spectroscopic coverage in the western part of the disk
we can investigate how the oxygen abundance varies along the arc-like region identified in Fig.~\ref{fig:f1}, 
designated as \mbox{\it arc A} hereafter, covering the restricted radial range $R/R_0$\,=\,0.54 to 0.59, and 
extending in position angle  between  230 and 290 degrees ($\sim$17~kpc in projected length), where we have a good amount of nebular spectra with \oiii\lin4363 detections, in order to test 
whether we could confirm the local azimuthal variation suggested by KG96.


\subsection {Abundance variations based on the direct method}\label{sec:Te}
There are 29 \hii\ regions with \Oiii$\lambda 4363$ line measurements in our sample, covering the radial range from $R/R_0=0.10$ (H493) to $R/R_0=1.25$ (SDH323). 
The resulting radial oxygen abundance gradient is presented in Fig.~\ref{fig:f4}. As the figure shows, the data can be well fitted by a weighted linear least-squares fit (the linear Pearson correlation coefficient is $-$0.90):

\begin{equation}\label{eq:radialte}
12+\log(O/H)_{T_e}= 8.73(\,\pm\,0.03)~ -~ 0.87(\,\pm\,0.04)~ R/R_0, 
\end{equation}

\noindent
which, unsurprisingly, is consistent with previous results by KBG03, $12+\log(O/H)_{T_e}= 8.76(\,\pm\,0.06) - 0.90(\,\pm\,0.08)~R/R_0$, and  B07, $12+\log(O/H)_{T_e}= 8.75(\,\pm\,0.05) - 0.90(\,\pm\,0.07)~R/R_0$.

\smallskip
Let us examine the abundance distribution along  arc A. As shown by the gray shadowed region in Fig.~\ref{fig:f4},  
this part of the galaxy contains objects whose O/H abundance ratios appear to cover a larger spread 
(at constant radius) than elsewhere. 
To clarify the situation, in Fig.~\ref{fig:f5} we zoom into this restricted radial range (top panel), and 
plot the oxygen abundance as a function of  position angle (bottom panel).
The red dotted lines mark the range in O/H corresponding to the change in galactocentric distance   from 0.54\,$R_0$ to 0.59\,$R_0$, as inferred from the gradient given by Eq~\ref{eq:radialte}. 

The \hii\ region H27 stands out with an abundance 12+log(O/H)\,=\,7.96$\,\pm\,$0.08,  
while the corresponding radial fit value at its galactocentric distance is 8.23$\,\pm\,$0.05. The difference between the measured and the fit value is 0.27 $\,\pm\,$ 0.09 dex. Therefore, H27 is a 3$\sigma$ outlier from the linear fit. An additional object observed by KBG03, H128, is also an outlier relative to the radial fit, with an observed value of 12+log(O/H)\,=\,8.45$\,\pm\,$0.04 dex and a fit value of 8.25$\,\pm\,$0.05 dex ($\sim$3.1$\sigma$). All other objects are consistent with the radial fit value for their galactocentric distances, within 3$\times$ their measurement uncertainties.

Like all other objects, H27 and H128 are located in the pure star-forming region of the BPT diagram (see Fig.~\ref{fig:f3}) and we find no evidence for the presence of other ionization mechanisms (e.g.~shocks) that could explain its peculiar oxygen abundance for its galactocentric distance. For example, we
find no significant enhancement in the strength of shock-sensitive  lines, such as \Sii, [O\,{\sc i}]. 
H27 lies within the region in which \citet{Franchetti:2012} made a spatially-detailed investigation of 
supernova remnant (SNR) candidates in M101 from archival Hubble Space Telescope images, and has been identified as a superbubble, rather than a SNR.

\smallskip
Analyzing the scatter relative to the radial abundance gradient is one approach to probe azimuthal variations of metals (\citealt{Bresolin:2011}). The rms scatter around the least-squares fit for our full sample of \hii\ regions (black solid line in Fig.~\ref{fig:f4}) is 0.11 dex.  For the 11 objects contained within the grey shadowed portion such scatter rises to 0.15 dex, while for the 18 remaining objects it is 0.08 dex. 

A scatter of 0.15 dex potentially represents a detection of local inhomogeneities in the abundance distribution.
In order to  test the significance of this result, we calculated the errors on the  scatter measured for these two data sets (inside and outside  the shadowed region in Fig.~\ref{fig:f4}), following a jackknife procedure \citep[][p.~46]{Lupton:1993}. Simply put, jackknife resamples a sample of size $N$, constructing $N$ subsamples of size $N-1$ by omitting an element of the sample in succession, and then estimates the variance of the scatter t$_{N}$ from

\begin{equation}
\sigma^{2}_{t_{N}}= \frac{N-1}{N}\sum_{i=1}^N (t_{N-1,i}-\bar{t}_{N-1})^2
\end{equation}

\noindent
where $\bar{t}_{N-1}$ is the average of t over the N subsamples. 

By applying this technique, the resulting scatters are 0.15$\,\pm\,$0.03 dex (objects within arc~A)
and 0.08$\,\pm\,$0.01 dex (objects outside arc~A), respectively. Thus, the significance of the difference is
only at the $\sim$2$\sigma$ level.

In summary, using the direct method on our enlarged sample of 29 \hii\ regions with \oiii\lin4363 detections, we derive a radial oxygen abundance gradient that is consistent with previous results. We detect two outliers from the mean radial gradient, H27 and H128, both of which with a significance at the 3$\sigma$ level. No evidence of shocked gas are found for H27. The objects located along arc A display an abundance scatter of 0.15$\,\pm\,$0.03 dex, which represents a marginal detection of a local azimuthal variation. 



\begin{figure*}
\centering
    \includegraphics{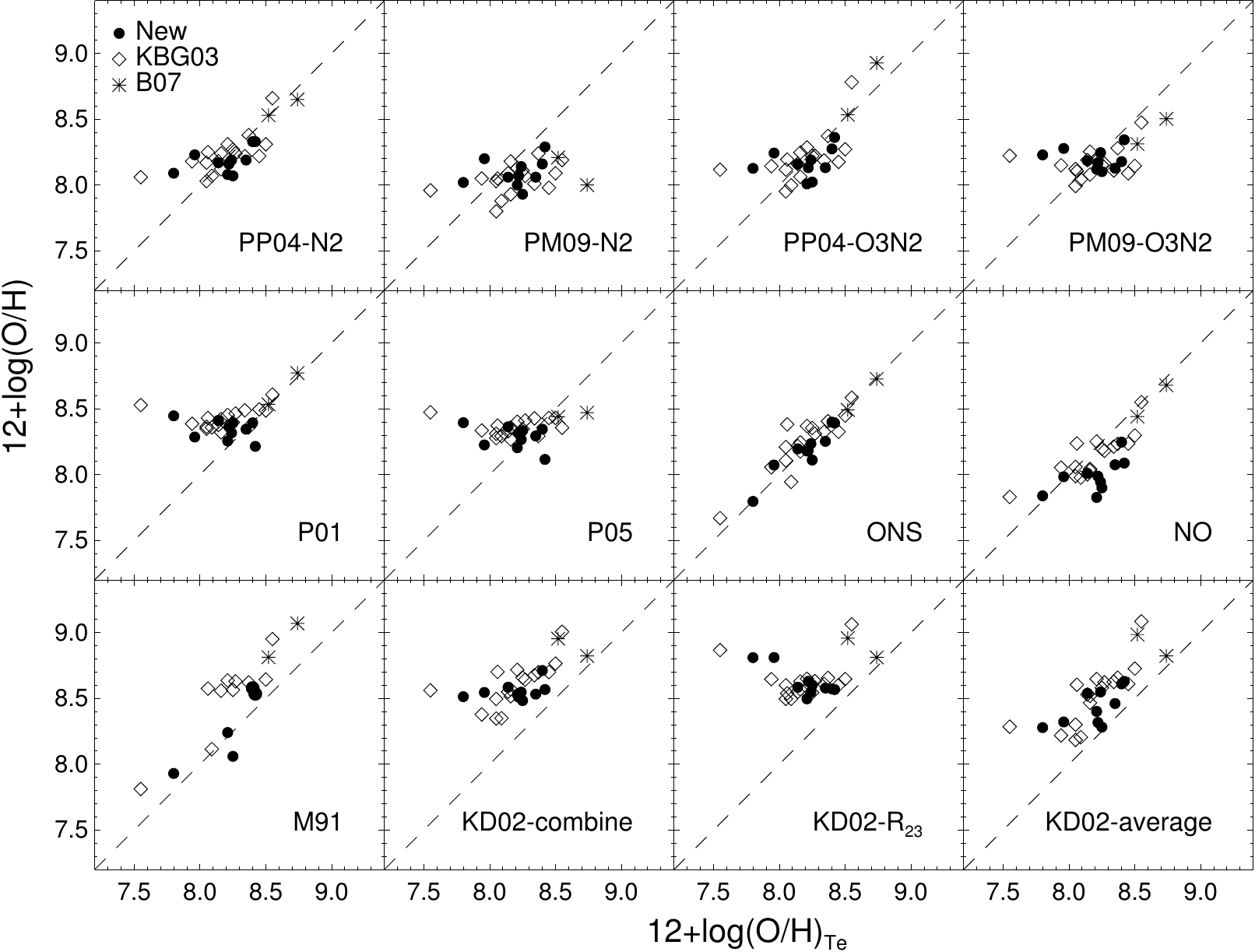}
    \caption{Comparison of oxygen abundances derived from different strong-line diagnostics with those obtained from the $T_e$ method. For the M91 method only 17 objects with a well-established $R_{23}$ branch attribution are plotted. \label{fig:f6}}
      
      \end{figure*}    

\subsection {Abundance variations based on strong-line methods}\label{sec:intro_sl}

\subsubsection{Comparison of different strong-line methods}
In addition to the \Oiii$\lambda4363$-based method, in order to take advantage of the 79 objects in our enlarged sample of \hii\ regions we also considered a number of strong-line methods, in which the strength of easily observed nebular emission lines can be used to calculate the nebular abundances. Among the  available methods making use primarily of O lines we considered the $R_{23}$ calibration by \citet[=\,M91]{McGaugh:1991} (in the analytical 
form provided by \citealt{Kobulnicky:1999}), and the P-method by \citet[][=\,P01]{Pilyugin:2001} and \citet[][=\,P05]{Pilyugin:2005}.
Other methods rely on the calibration of the strength of other emission lines, typically the N lines, in terms of the O abundance: \citet[][PP04-N2 and PP04-O3N2]{Pettini:2004}, \citet[][PM09-N2 and PM09-O3N2]{Perez-Montero:2009a}, \citet[][ONS]{Pilyugin:2010},  \citet[][KD02]{Kewley:2002} and the empirical calibration of \Nii/\Oii\ from \citet[][NO]{Bresolin:2007}. These methods are summarized in Table~\ref{table:calibrations} for clarity. 
They are somewhat arbitrarily divided in `empirically' and 'theoretically' calibrated methods, to reflect the
fact that the calibration in terms of O/H abundance ratio can be obtained from samples of \hii\ regions with available direct abundances or from grids of photoionization models, respectively.

\smallskip
For our purposes, we were interested to test how the strong-line abundances compared with the abundances from the direct method.
We show in Fig.~\ref{fig:f6} the comparison between \hii\ region abundances obtained from the $T_e$ method (along the x-axis) and those obtained from the various strong-line diagnostics. 
Since similar comparisons have appeared frequently in the literature in the past decade, presenting comparable results, we keep our discussion limited to the essential. For further details, the reader is referred to, for example: \citet{Kennicutt:2003}, \citet{Bresolin:2009a}, \citet{Kewley:2008}, \citet{Perez-Montero:2005}, and \citet{Yin:2007}.

\medskip
\noindent
{\it Empirically calibrated methods} --
The first two rows of Fig.~\ref{fig:f6} refer to empirically calibrated diagnostics.
A quick look at the figure shows that despite the fact that, by construction, these methods should generally agree with the direct method, the level of  agreement varies greatly, at least for the sample of \hii\ regions we are considering. In some cases this is due to the fact that the O/H range in which
the diagnostics were calibrated is narrower than the full range of abundances available in M101 (which is unusually large). Secondly,
possible differences in physical conditions (e.g.~ionization parameter) between the calibrating sample and our sample can translate into the variations we observe.
With these considerations in mind, we point out the following points:

\begin{itemize}

\item PP04-N2 provides a relatively good match to the direct abundances (with a rms scatter of 0.11 dex), but large discrepancies arise at low metallicity, 12+log(O/H)\,$<$\,8.0. Similarly for PP04-O3N2, but the scatter 
is worse (0.17~dex).

\item PM09-N2 tends to systematically underestimate the abundances, compared to the direct method. PM09-O3N2 yields similar results.

\item the P01 calibration was derived from  nebular data with 12+log(O/H) $>$ 8.2, which is therefore the applicable range of this method. Accordingly, it is not surprising to see large discrepancies in the low metallicity regime. 

\item P05 is a revised version of P01,  providing  calibrations for both the upper and the lower branch of the relation between oxygen abundance and $R_{23}$.  Here, we only adopted the calibration  for the upper branch,
but even at high metallicity the diagnostic underestimates the $T_e$-based abundances.

\item the ONS calibration provides abundances that are comparable to those from the $T_e$ method across the whole metallicity range, with a rms scatter of 0.09 dex.

\item the NO calibration seems to largely agree with the $T_e$ method, although the scatter of the data points is larger than in the case of the ONS method.

\end{itemize}

\begin{deluxetable*}{lcccc}
\tabletypesize{\scriptsize}
\tablecaption{Summary of  nebular abundance diagnostics adopted\label{table:calibrations}}
\tablehead{
\colhead{Number} & 
\colhead{Method} &
\colhead{Emission lines involved} & 
\colhead{Calibration class} &
\colhead{References}}
\startdata 
1& $T_{e}$         	    & \Oiii$\lambda4363$, \Oiii$\lambda$4959,5007  & Direct       &     	 \citet{Aller:1984,Stasinska:2005}\\   
2&  PP04-N2     	    & \Nii/H$\alpha$       						& Empirical &		\citet{Pettini:2004}\\
3&  PM09-N2      	    & \Nii/H$\alpha$, \Nii/\Oii			     		& Empirical &		\citet{Perez-Montero:2009a}\\
4& PP04-O3N2		    & \Nii/H$\alpha$, \Oiii/H$\beta$     				& Empirical &		\citet{Pettini:2004}\\
5& PM09-O3N2 	    & \Nii/H$\alpha$, \Oiii/H$\beta$,\Nii/\Oii     		& Empirical &		\citet{Perez-Montero:2009a}\\
6& P01\tablenotemark{a} & R$_{23}$, \Oiii/\Oii						& Empirical &		\citet{Pilyugin:2001}\\
7& P05\tablenotemark{a}  & R$_{23}$, \Oiii/\Oii					& Empirical &		\citet{Pilyugin:2005}\\
8& ONS          		    & \Sii/\Oii, \Nii/\Oii, \Oiii/H$\beta$        		         & Empirical &  		  \citet{Pilyugin:2010}\\			
9& NO               		   &  \Nii/\Oii								& Empirical &  		  \citet{Bresolin:2007}\\
10& M91        		   & R$_{23}$, \Oiii/\Oii 						& Theoretical & 	\citet{McGaugh:1991,Kobulnicky:1999}\\
11& KD02-combine\tablenotemark{b}  	 & \Nii/\Oii, R$_{23}$, \Oiii/\Oii, \Nii/\Sii & Theoretical & 	\citet{Kewley:2002}\\
12& KD02-R$_{23}$\tablenotemark{c}    & \Nii/\Oii, R$_{23}$, \Oiii/\Oii	& Theoretical & 		 \citet{Kewley:2002}\\
13& KD02-average\tablenotemark{d}    & \Nii/\Oii, R$_{23}$, \Oiii/\Oii	& Theoretical &			 \citet{Kewley:2002}
\enddata

\tablenotetext{a}{P $\equiv$ \Oiii$\lambda$$\lambda$4959,5007/(\Oii$\lambda$3727+\Oiii$\lambda$$\lambda$4959,5007).}
\tablenotetext{b}{diagnostic recommended for an optimized abundance determination in KD02.} 
\tablenotetext{c}{the $R_{23}$ method in KD02.}
\tablenotetext{d} {the `combine average' method by KD02, which is the average of five  independent strong-line methods.}

\end{deluxetable*}


\medskip
\noindent
{\it Theoretically calibrated methods} -- The last row of Fig.~\ref{fig:f6} presents the abundances obtained from strong-line methods that are calibrated from theoretical models. It is well-known that they provide systematically higher abundances than the $T_e$ method (e.g.~\citealt{Bresolin:2009a}).
`KD02-combine' represents the so-called `combined' diagnostic of KD02, recommended as an optimized abundance determination by these authors, and which can be used over the range of abundances from 12+log(O/H) $=$ 8.2 up to 9.4,  with a minimized scatter and showing no systematic offset compared with `KD02-average'. 
`KD02-R$_{23}$' is the R$_{23}$ method calibrated from the models of KD02: compared to the $T_e$ method it shows a large scatter both at  high  and low metallicities. 
`KD02-average' is the average of five independent strong-line methods. As we can see in Fig.~\ref{fig:f6}, it  overestimates the $T_e$-based abundances, 
but the relative offset appears unchanged across the full metallicity range.

  \begin{figure}
\centering
    \includegraphics[width=1.\columnwidth]{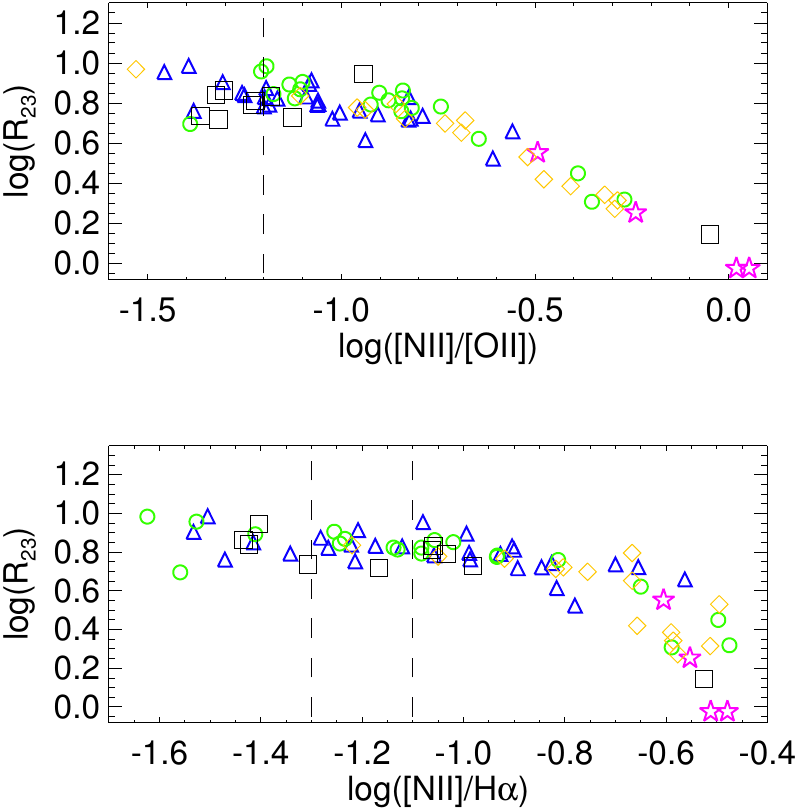}
    \caption{{\it Top:} Observed relationship between the \Nii$\lambda6583$/\Oii$\lambda3727$ line ratio and the log($R_{23}$) indicator. {\it Bottom:} Observed relationship between the \Nii$\lambda6583$/H$\alpha$  line ratio and the log($R_{23}$) indicator.  The dashed lines indicate the criteria to distinguish between the $R_{23}$ upper and lower branches according to \citet{Kewley:2008}. Symbols as in Fig.~\ref{fig:f1}.\label{fig:f7}}
       
      \end{figure}

     \begin{figure*}
\centering
    \includegraphics[width=1.7\columnwidth]{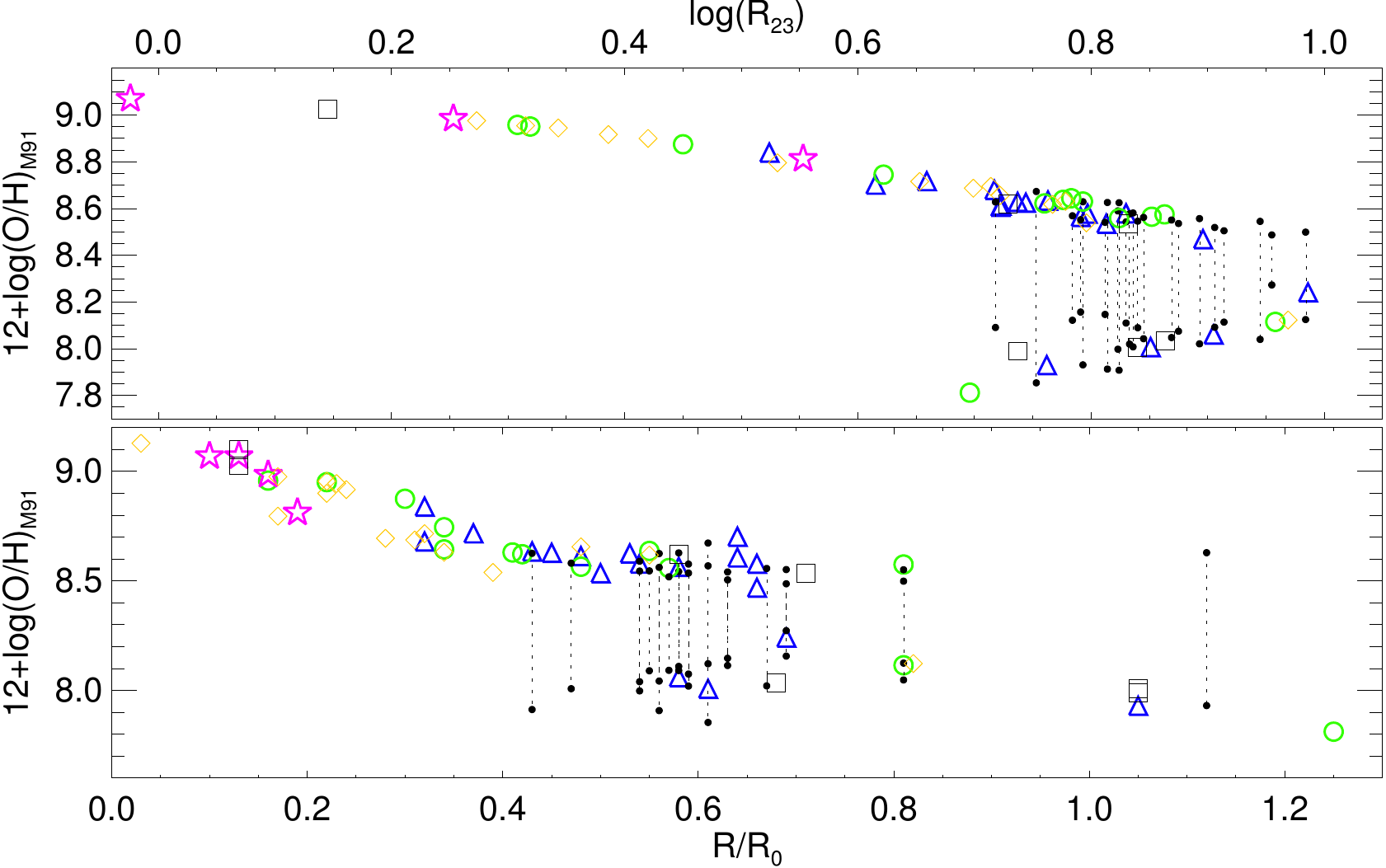}
    \caption{{\it Top:} oxygen abundances from $R_{23}$ as a function of log($R_{23}$). {\it Bottom:} radial oxygen abundance distribution based on the M91 calibration of the $R_{23}$ diagnostic. There are 22 objects (black filled circles) whose branches could not be univocally established and both upper and lower branch O/H values are shown as black dots  connected with dot lines. Symbols as in Fig.~\ref{fig:f1}.\label{fig:f8}}
       
      \end{figure*}

Finally, M91 suffers from the well-known double-valued nature of the  $R_{23}$
diagnostic.  
Additional line ratios and/or an initial guess of the abundance are required to choose the appropriate branch.
We attempted to break the $R_{23}$ degeneracy by using the \Nii/\Oii\  and  \Nii/H$\alpha$ ratios, following the criteria found   in \citet{Kewley:2008}. According to these authors,  the separation  between the $R_{23}$ upper and lower branches   occurs at log(\Nii/\Oii)\,=\,$-1.2$ and $-1.3$\,$\le$\,log(\Nii/\Ha)\,$\le$\,$-1.1$. These boundary values are marked by dashed lines in Fig.~\ref{fig:f7}, where we plot the trends of both 
\Nii/\Oii\ and \Nii/H$\alpha$ with $R_{23}$. Only when both these criteria were simultaneously satisfied we assigned the objects to the appropriate branch. 

However, there is a large number of 22 objects in our sample whose branches could not be firmly established, or remain dubious or contradictory, based on these criteria. These objects are represented by black filled circles connected by dotted lines in Fig.~\ref{fig:f8}, where we show how the M91-derived abundances change as a function of $R_{23}$ parameter {\it (top panel)} and galactocentric distance {\it (bottom panel)}.
As the bottom panel of this figure
shows dramatically, the $R_{23}$ branch choice can yield very different conclusions about the radial trend of the oxygen abundance. This can make double-valued diagnostics like $R_{23}$ 
inappropriate to study radial abundance gradients in galaxies, and as a minimum the results should be
checked with additional diagnostics.

\smallskip
In summary, the ONS calibration gives the best agreement with the $T_e$ method for the oxygen abundances of the M101 \hii\ region sample among the strong-line diagnostics considered in our study. Most of the remaining empirically-calibrated strong-line methods diverge from the $T_e$ method at low metallicities. The theoretically-calibrated diagnostics tend to systematically overestimate the abundances, compared to the $T_e$ method.
The excellent agreement of the ONS method with the $T_e$ method is not too surprising, given that it was specifically calibrated using \hii\ regions with well-measured $T_e$-based abundances (but so where the others). To further test its agreement with the $T_e$ method we applied the ONS method to the \hii\ region sample in the galaxy NGC~300 presented by \citet{Bresolin:2009a}, who found good agreement between the nebular abundances derived with the $T_e$ method and the metallicities of blue supergiant stars (\citealt{Kudritzki:2008}).
The abundance gradient from the ONS method is found to be represented by the following linear fit:
12+log(O/H)\,=\,8.49($\,\pm\,$0.02)~$-$~0.33($\,\pm\,$0.04)~$R/R_0$. This agrees well with the fit obtained from the $T_e$ method, 12+log(O/H)\,=\,8.57($\,\pm\,$0.02)~$-$~0.41($\,\pm\,$0.03)~$R/R_0$, and from the blue supergiants, 12+log(O/H)\,=\,8.59($\,\pm\,$0.05)~$-$~0.43($\,\pm\,$0.06)~$R/R_0$.


\subsubsection {Abundance variations based on the ONS diagnostic}\label{s:ons}

      \begin{figure*}
\centering
    \includegraphics{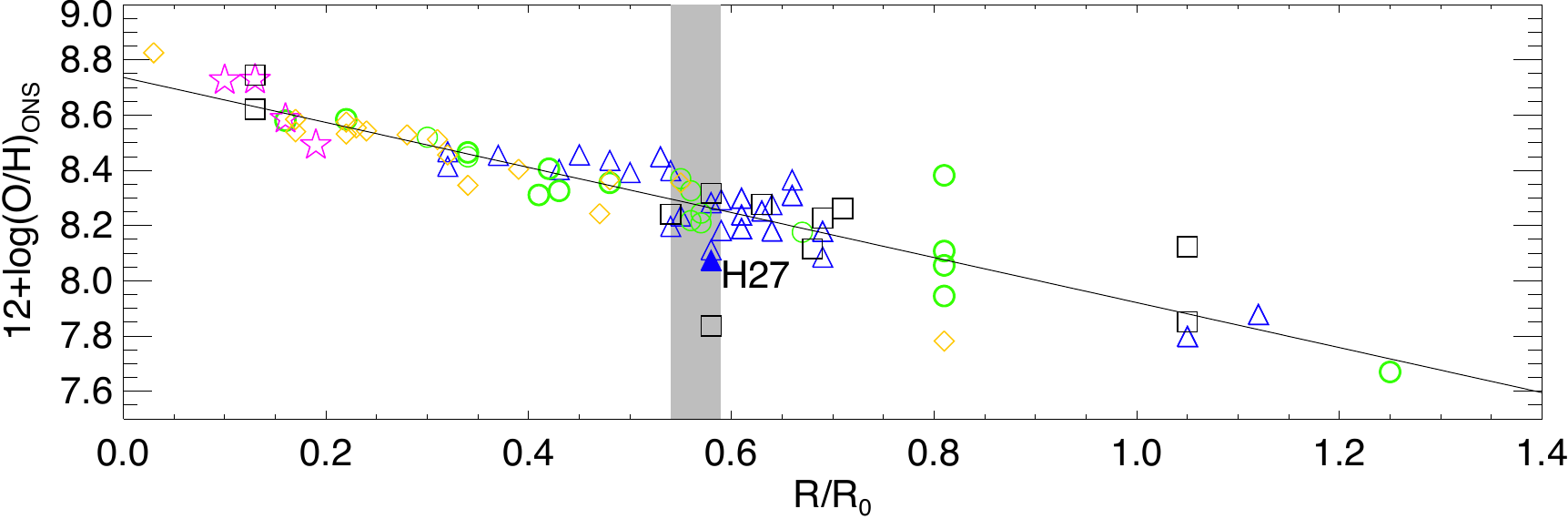}
    \caption{Radial oxygen abundance  gradient from the ONS method. The solid line is the linear least-squares fit including 74 \hii\ regions. H27 is marked by the filled triangle. The grey shadowed region is the same  as in Fig.~\ref{fig:f4}. Symbols as in Fig.~\ref{fig:f1}.
      \label{fig:f9} }
      \end{figure*}
 
\begin{figure}
\centering
    \includegraphics[width=0.9\columnwidth]{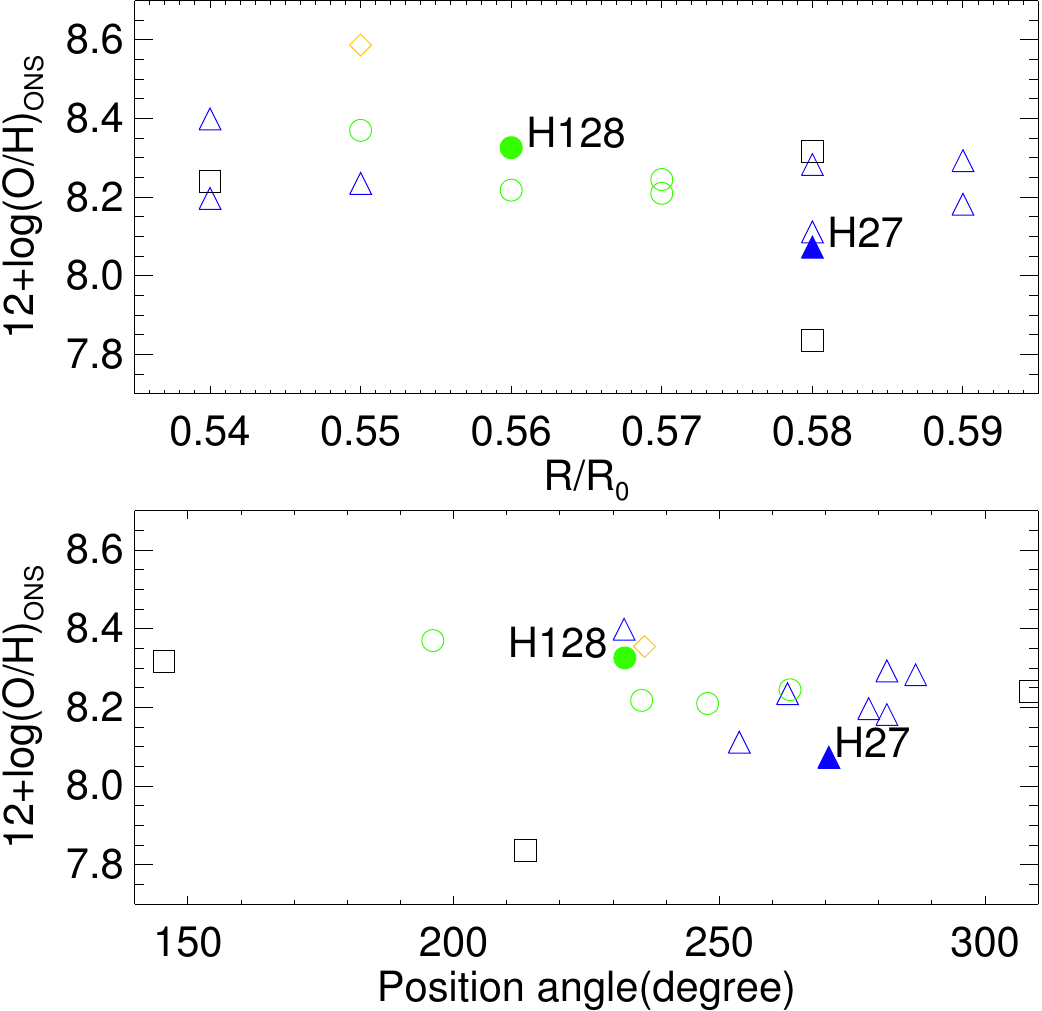}
    \caption{Detailed view of the  shadowed region in Fig.~\ref{fig:f9}. 
    The upper panel shows the relation between  oxygen abundance and  galactocentric distance in the range  $R/R_0$\,=\,0.54 to 0.59. The bottom panel is the relation of oxygen abundance with  position angle. The filled circle and triangle represent H128 and H27, respectively. Symbols as in Fig.~\ref{fig:f1}.\label{fig:f10} } 
      
      \end{figure}

In order to expand our analysis of possible  spatial variations of the chemical abundances  presented in Sect.~\ref{sec:Te}, we used the oxygen abundances derived from the ONS method, which shows the best agreement with the 
$T_e$ method, for the enlarged sample of \hii\ regions. 
However, adopting other strong-line diagnostics would yield the same conclusions concerning azimuthal variations in the inner disk.

The radial oxygen abundance  gradient thus derived  is shown in Fig.~\ref{fig:f9}. The solid line represents the least-square fit to the data. 
We note that there are five objects at $R/R_0$\,=\,0.81 displaying a large abundance scatter. These are five individual emission knots located within the supergiant \hii\ region NGC~5471 (see KG96). 
At least two of them have been associated with SNRs or the presence of high-velocity gas \citep{Skillman:1985,Chu:1986,Kennicutt:1996}. Peculiar ionization conditions in these knots could in principle affect the chemical abundance determination (but we note that the $T_e$-based oxygen abundances do not appear to be  affected, and are quite homogeneous between the different knots).
Thus, the apparently large abundance scatter between these data points, as seen in Fig.~\ref{fig:f9}, is unlikely to result from a real oxygen abundance inhomogeneity in this region (an interpretation supported by the $T_e$-based O/H measurements; see KBG03 for a discussion of the N/O ratio), but rather 
from a breakdown of the ONS diagnostic in this nebular complex.

The linear fit to the radial abundance gradient is not significantly affected by the inclusion of the knots in NGC~5471. We obtain: 

\begin{equation}
12+\log(O/H)_{ONS}= 8.74(\,\pm\,0.03)-0.82(\,\pm\,0.05)~R/R_0
\end{equation}

\noindent
with a linear Pearson correlation coefficient of $-$0.90.
The intercept and the slope of the regression are consistent with the  result derived from the direct method (Eq.~\ref{eq:radialte}).

As done earlier, in Fig.~\ref{fig:f10} (top panel) we zoom into the radial range $R/R_0=0.54$ to 0.59 (highlighted in gray in Fig.~\ref{fig:f9}), and plot
the oxygen abundance as a function of  position angle (bottom panel). \hii\ regions H27 (filled blue triangle) and H128 (filled green circle), which were found to be possible outliers from the radial abundance gradient obtained from the direct method, are not so using the ONS method. However, a new object, corresponding to slit~6 in \citet{van-Zee:1998a}, is now found to have quite a low O/H abundance ratio, \oh\,=\,7.84. This can be explained by the fact that the criterion used in the ONS method to assign an \hii\ region to a particular excitation class fails for this particular nebula. The criterion, based on the \nii/\sii\ line ratio, assigns this object to the `hot' class, but if we calculate its O/H abundance as if it were a 'warm' \hii\ region, its O/H ratio would increase by 0.3 dex, which would thus remove the systematic abundance offset relative to the other nebulae with similar galactocentric radius. The problem of misclassification in the ONS method has already been discussed by \citet{Bresolin:2012}. In addition, for the specific case of M101 \citet{Pilyugin:2010} recommended to use the `'hot' classification only for objects lying at $R/R_0 > 0.90$.

The rms scatter around the least-square fit for  74 \hii\ regions (the enlarged sample with the five knots in NGC~5471 removed)  is  0.10 dex. For the 17 objects in the grey shadowed region of Fig.~\ref{fig:f9}
the scatter is 0.14 dex, and for all the remaining objects it is 0.09 dex. However, if we focus on the 13 objects along arc A {\it only} we find that the scatter is comparable to what we find for  the remaining objects ($\sim$0.10 dex).
In other words, based on the ONS method, we find no evidence for local inhomogeneities of the oxygen abundance along  arc~A. 




\section{A global asymmetry in the oxygen abundance?}\label{s:halves}

  \begin{figure}
\centering
        \includegraphics[width=1.\columnwidth]{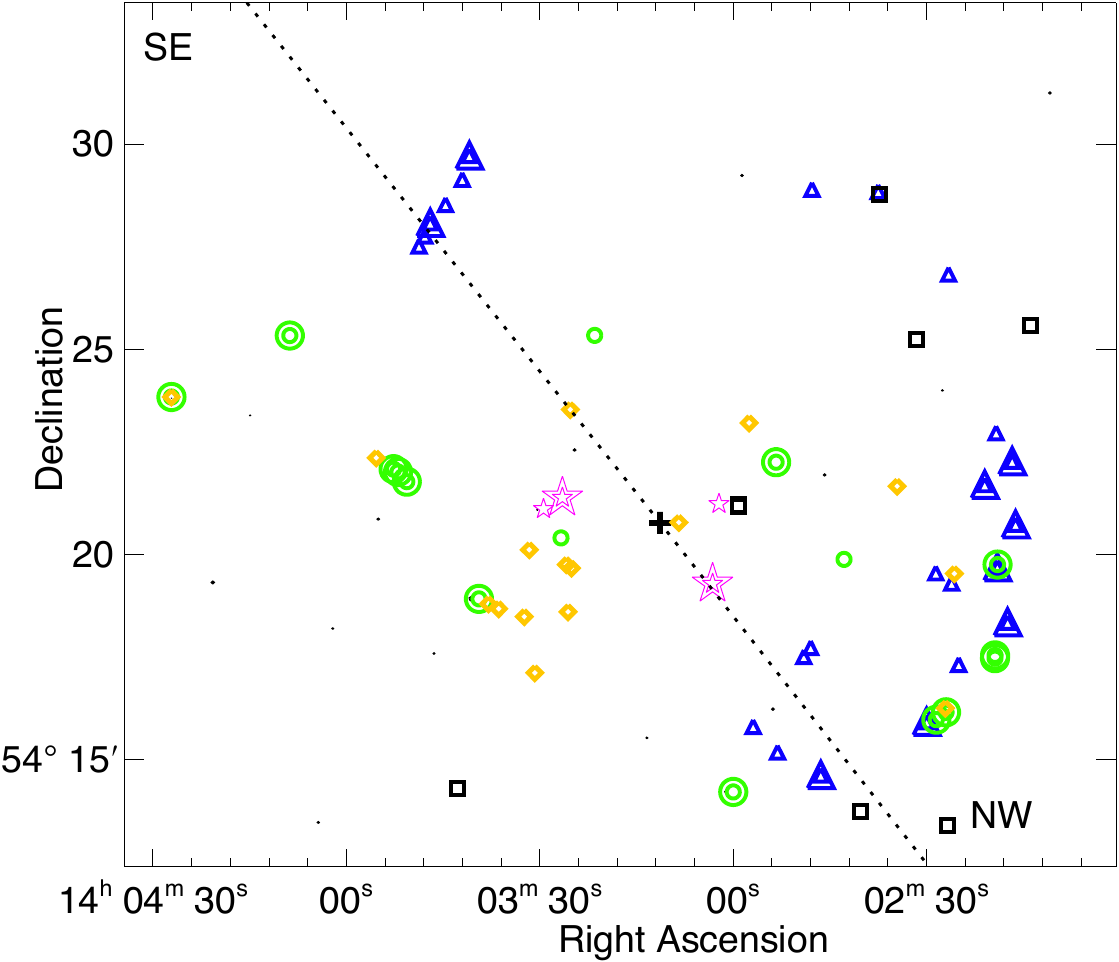}

    \caption{Spatial distribution of the \hii\ regions in our sample in the plane of the sky. Our sample is geometrically divided into the `SE' (with position angle relative to the galaxy center between 37 and 217 deg) and  `NW' part (complementary values of position angle). Symbols as in Fig.~\ref{fig:f1}.
    \label{fig:f11}}
       
      \end{figure}
     
\begin{figure}
\centering
        \includegraphics[width=1\columnwidth]{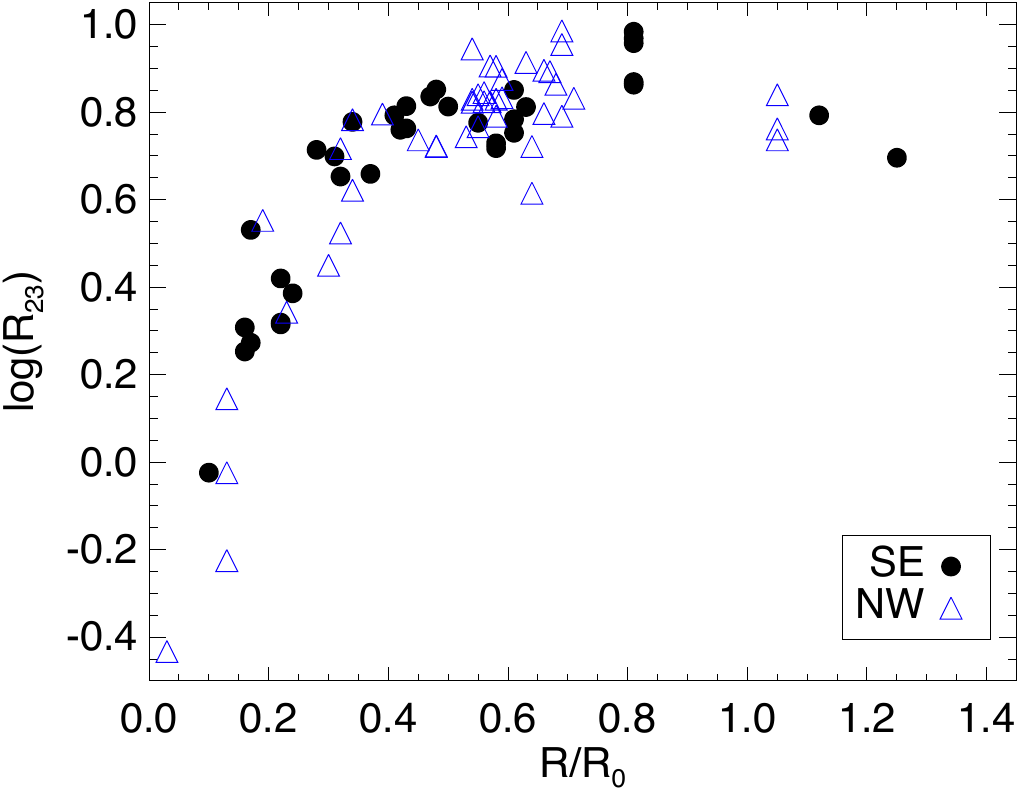}

    \caption{Radial dependence of the abundance parameter $R_{23}$. Filled circles$-$\hii\ regions in the SE, open triangles$-$\hii\ regions in the NW. \label{fig:f12}} 
     
      \end{figure}

It is interesting to test not only for local chemical inhomogeneities, as done in the previous sections, but also for 
large-scale azimuthal variations in the disk of M101.
Is it true that the metal abundance distribution in the disk is azimuthally axisymmetric, as usually assumed in chemical evolution models? To start to answer this question in the case of M101, further examination of the two dimensional distribution of the oxygen abundance is required. 

As a simple test, we divided the galaxy into two halves with respect to the major axis: a SE part (position angle: 37 to 217 degrees, including 37 objects) and a NW part (complementary position angles, with 42 objects), as shown in Fig.~\ref{fig:f11}. 
KG96 suggested that objects in the SE  have lower oxygen abundances compared to \hii\ region in the NW. This result was obtained from the use of the $R_{23}$ parameter.
We plotted the radial dependence of $R_{23}$ in M101 in Fig.~\ref{fig:f12} in the same way KG96 did, but including about twice the number of objects  (79 versus 37). The distribution of data points is virtually 
the same for the two subsamples, and thus we do not confirm the non-axisymmetric distribution suggested by KG96 considering the $R_{23}$ diagnostic alone.

\begin{figure*}
\centering
                \includegraphics[width=0.9\columnwidth]{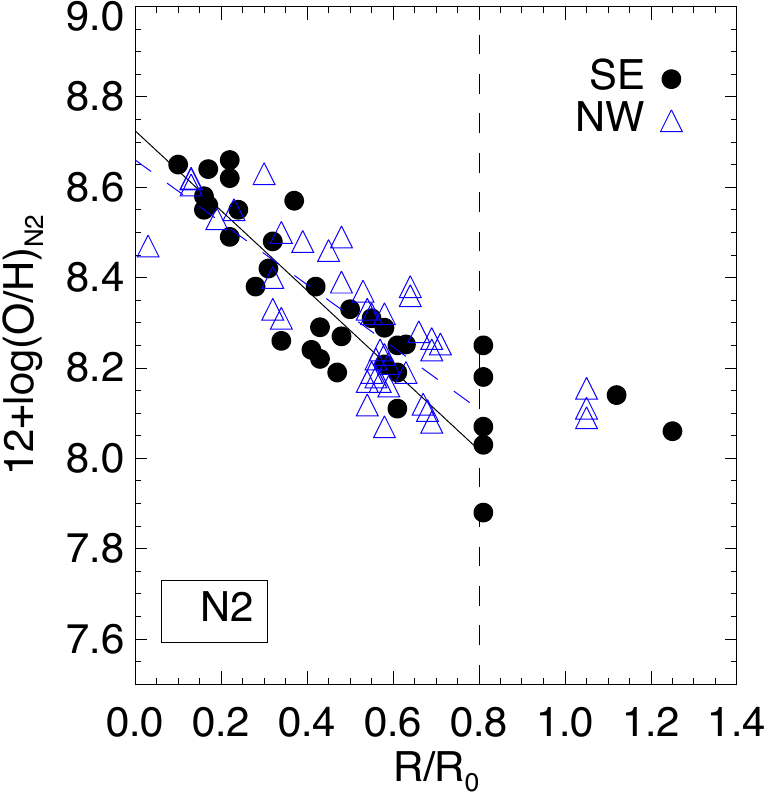}
                 \includegraphics[width=0.9\columnwidth]{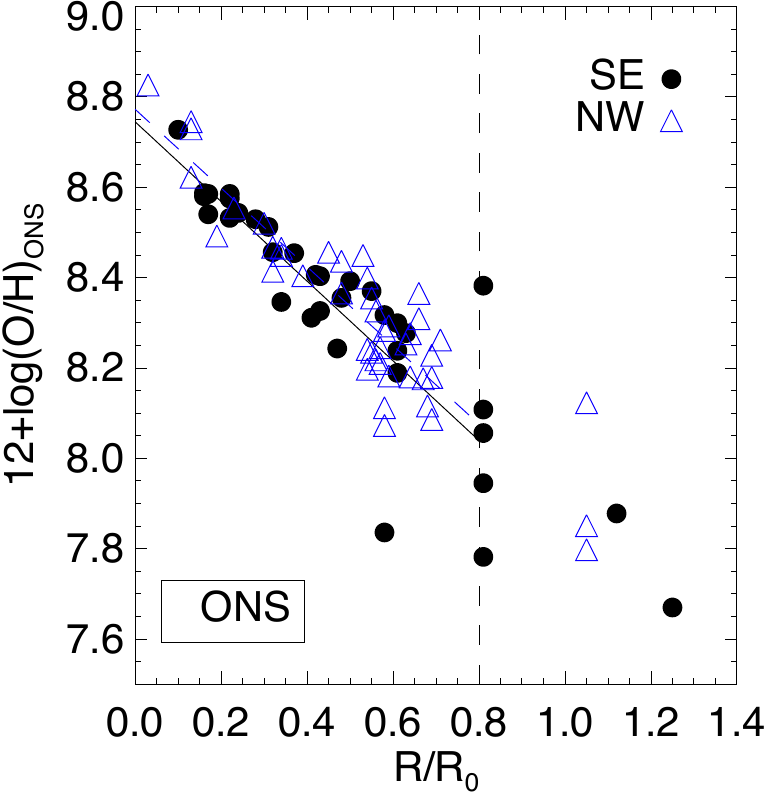}
       \includegraphics[width=0.9\columnwidth]{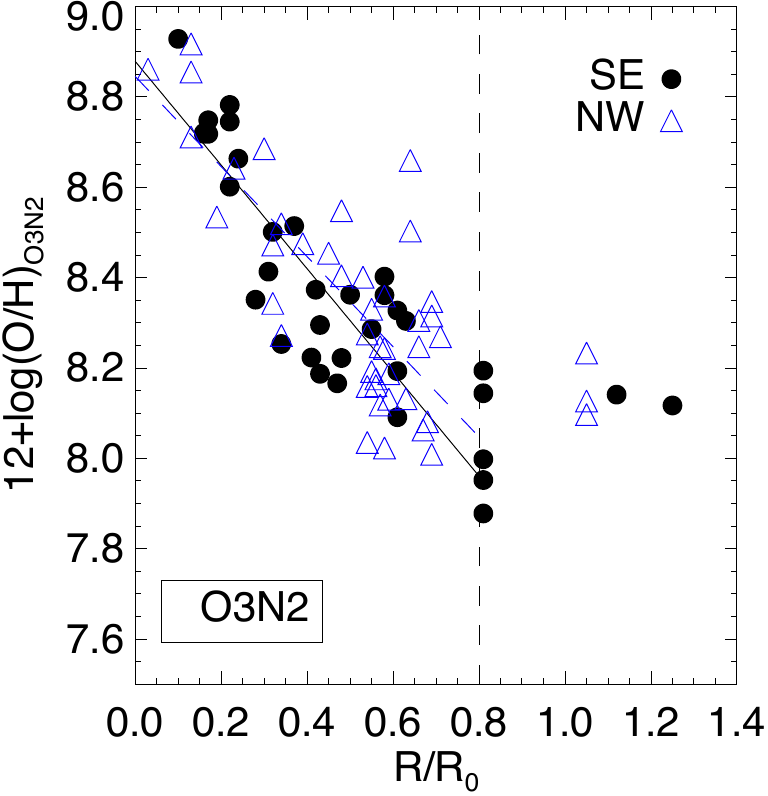}
          \includegraphics[width=0.9\columnwidth]{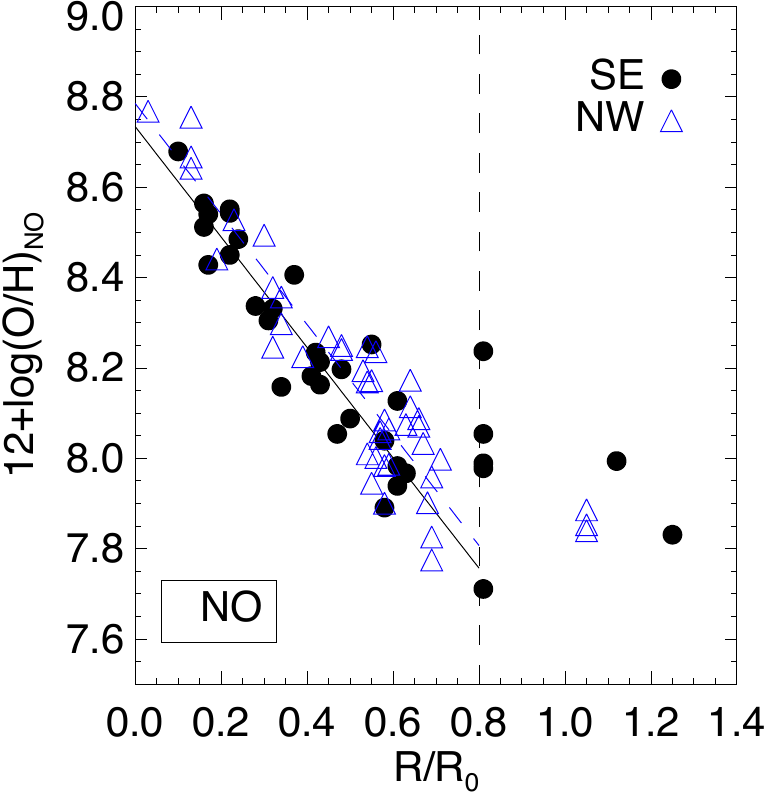}   
                   \includegraphics[width=0.9\columnwidth]{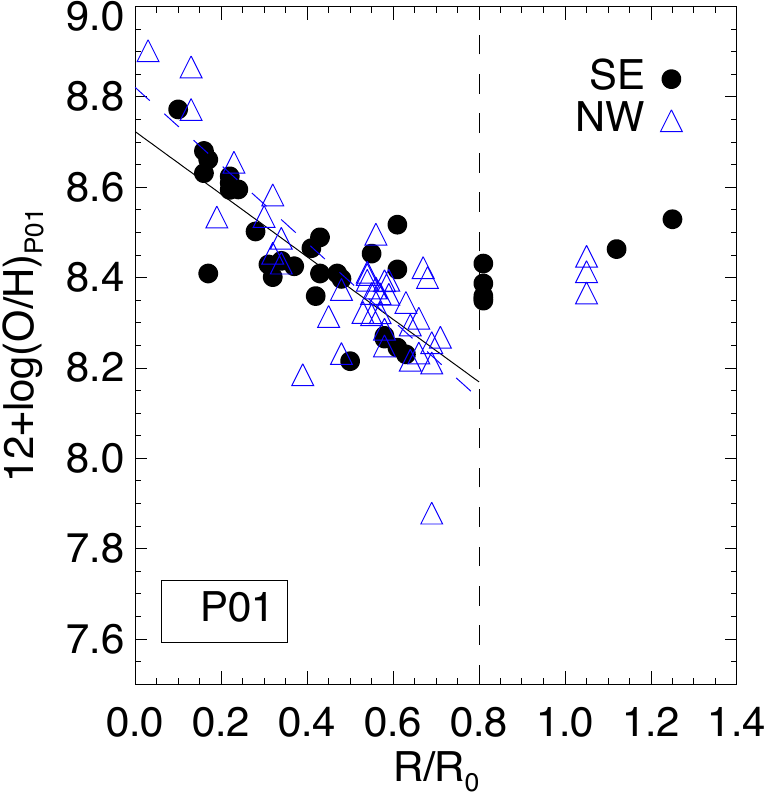}
             \includegraphics[width=0.9\columnwidth]{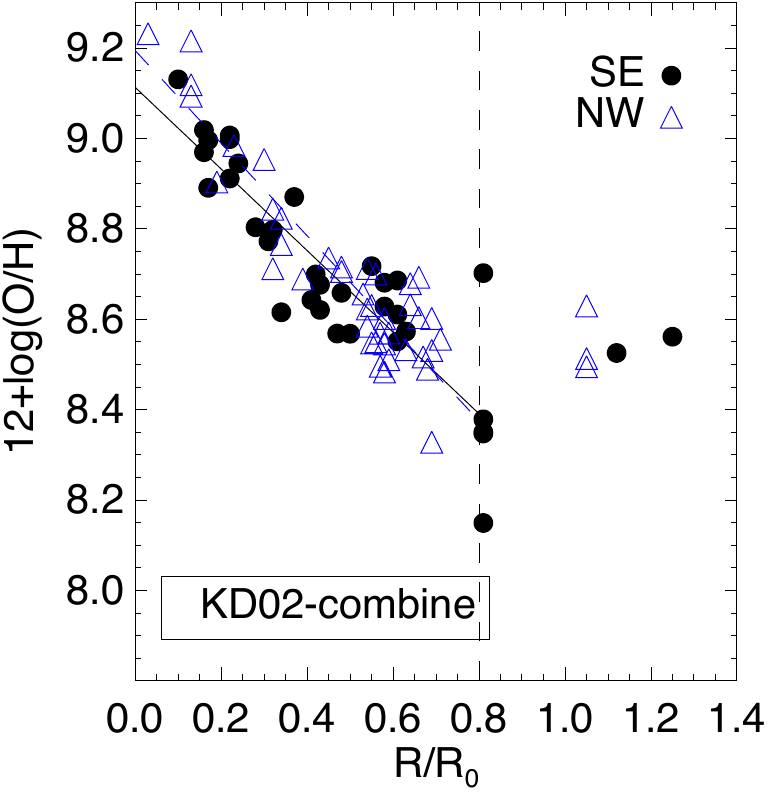}
             
\caption{Radial oxygen abundance gradients in the SE and NW sections of M101 based on six different strong-line methods.
The straight lines show the weighted linear regressions to the two samples, limited to data points located at $R<0.8~R_0$ (marked by the vertical dashed line).
\label{fig:f13} }
      
\end{figure*}

As a further test, we used six different strong-line methods to calculate the oxygen abundances and fitted the radial distribution with linear functions of the form 12+log(O/H)\,=\,$a$~+~$b$$\times$($R/R_0$). 
Since we find (see Fig.~\ref{fig:f13}) that different abundance diagnostics yield qualitatively different radial trends in the outer disk, only objects with $R/R_0$$<$0.8 were included in the fit.
In Table~\ref{table:2sectors} we summarize, for each diagnostic, the linear fit parameters $a$ and $b$ and their uncertainties, the Pearson correlation coefficients, the rms scatter of the  fit  and the characteristic oxygen abundances measured  at $R$\,=\,0.4\,$R_0$, which correlate with the integrated galaxy metallicity \citep{Zaritsky:1994,Moustakas:2006}. Fig.~\ref{fig:f13} illustrates the results of the fitting procedure for the two \hii\ region subsamples (SE and NW) for each diagnostic. 
Table~\ref{table:2sectors} shows that the linear radial gradient parameters (e.g.~slope and zero point) are highly dependent on the abundance diagnostic adopted, which is 
not surprising given the different behaviors found among the strong-line methods when comparing with the
$T_e$ method.
The possible breakdown of many of the diagnostics at low metallicities  is evident in Fig.~\ref{fig:f13}, which shows that in several cases (e.g.~for P01 and KD02-combine) the radial abundance gradient reaches a minimum abundance at $\sim$0.8~$R_0$, and then rises again in the outer disk. This situation  has been more thoroughly discussed by \citet{Pilyugin:2003}. 
If we believe that these results are unphysical, i.e.~that 
the actual galactic gradient does not `turn over' at a certain galactocentric distance, we need to discard 
the abundances inferred from the strong-line diagnostics that are affected by this issue (we stress that  
the direct abundance are not affected). 
However, we point out that in recent years several reports of metallicity gradients becoming flat or turning over at large galactocentric radii in the Milky Way and other galaxies have been made (e.g.~\citealt{Worthey:2005, Yong:2012, Bresolin:2009}). The radius where such a break appears to occur in M101 agrees well with the results from the recent work of \citet{Scarano:2013}, who
noted a good correlation between the radial positions of the break and corotation radii for a sample of spiral galaxies. Our Fig.~\ref{fig:f13} shows, however, that the various abundance diagnostics we considered still leave some ambiguity concerning both the presence of the break and  the gradient slope in the {\it outer disk} (this problem is absent in other outer disk nebular abundance studies, e.g.~\citealt{Bresolin:2009, Bresolin:2012}).
Since the main goal of this work is to investigate the possibility of azimuthal variations in the {\it inner disk}, where the metallicity trends are less ambiguous, we decided 
to limit our analysis to $R/R_0$\,$<$\,0.8.

Doing so, the differences in slope and  intercept of the linear regression between the SE and the NW portions of the disk are found to be smaller than the corresponding uncertainties, confirming the absence of obvious large-scale abundance variations in the azimuthal direction. A much more extensive spectroscopic coverage of M101 would be required to test for the presence of azimuthal variations among smaller sections (e.g.~quadrants) of the disk.

\begin{deluxetable*}{lccccccc}
\tablecaption{Parameters of the separate linear fits to the radial gradients \label{table:2sectors}}
\tablehead{
\colhead{Region} & 
\colhead{$a$} &
\colhead{$b$} & 
\colhead{$\sigma(a)$} &
\colhead{$\sigma(b)$} &
\colhead{correlation coeff.} &
\colhead{scatter (rms)} &
\colhead{12+log(O/H)$_{R=0.4R_0}$}
}

\startdata 
%
\sidehead{{\bf \footnotesize PP04-N2} }		
ALL &      8.68&    $-$0.75&    0.03&     0.06&	$-$0.83&     0.10&  8.38\\
SE&      8.72&    $-$0.88&    0.04&     0.10&	$-$0.87&     0.08& 8.37\\
NW&     8.66&     $-$0.69&    0.04&     0.09&	$-$0.79&     0.10&  8.38\\

\sidehead{{\bf \footnotesize PP04-O3N2} }		
ALL &      8.85&    $-$1.03&   0.05&    0.09&$-$0.80&     0.14&  8.44\\
SE&      8.88&    $-$1.15&   0.06&     0.15&$-$0.84&     0.13& 8.42\\
NW&     8.84&    $-$1.00&   0.07&     0.13&$-$0.77&     0.15& 8.45\\

\sidehead{{\bf \footnotesize P01} }		

ALL &      8.77&    $-$0.78&   0.03&    0.07&$-$0.81&      0.10&  8.46\\
SE&      8.72&    $-$0.69&    0.04&     0.10&$-$0.79&     0.09& 8.45\\
NW&     8.82&    $-$0.86&    0.05&     0.09&$-$0.82&     0.11& 8.48\\

\sidehead{{\bf \footnotesize ONS} }		

ALL &      8.75&    $-$0.85&   0.03&     0.06&$-$0.87&    0.09& 8.41\\
SE&      8.74&    $-$0.88&   0.05&     0.11&$-$0.84&    0.10&8.39\\
NW&     8.77&    $-$0.87&   0.04&     0.07&$-$0.89&     0.09&8.42\\

\sidehead{{\bf \footnotesize NO} }		

ALL &      8.75&     $-$1.18&   0.03&     0.06&$-$0.92&    0.09& 8.27\\
SE&      8.73&     $-$1.22&    0.04&    0.10&$-$0.93&    0.08&8.25\\
NW&     8.78&     $-$1.22&   0.04&     0.08&$-$0.93&    0.09&8.30\\

\sidehead{{\bf \footnotesize KD02-combine} }		

ALL &      9.15&    $-$0.96&    0.03&     0.06&$-$0.90&    0.08&8.77\\
SE&      9.11&    $-$0.90&    0.04&     0.10&$-$0.88&    0.08&8.75\\
NW&     9.19&     $-$1.02&    0.04&     0.07&$-$0.92&    0.08&8.78
\enddata
\tablecomments{Linear fits $a+b\times$($R/R_0$) including only  objects with $R/R_0$ $\leq$ 0.8}

\end{deluxetable*}

\section{Summary and conclusions}\label{s:conclusion}
Using a data sample of 79 \hii\ regions, with 28 from our new observations (yielding 10 new detections of the \Oiii$\lambda4363$ line) and the rest from additional sources in the literature, we have  obtained and analyzed the radial oxygen abundance   gradient in M101 based on the direct method (\Oiii$\lambda$4363-based), as well as various strong-line diagnostics. 
We found an exponential abundance profile with a gradient of $-$0.87$\,\pm\,$0.04 dex R$_0$$^{-1}$ and a central abundance of 12+log(O/H)$=$8.73$\,\pm\,$0.03, from the direct oxygen abundance measurements. The scatter in the radial abundance gradient along arc~A, located in the western disk of M101, is 0.15$\,\pm\,$0.03 dex, while for the remaining \hii\ regions of the M101 disk the scatter is 0.08$\,\pm\,$0.01 dex. 
In the same section of the galaxy we  found that one \hii\ region, H27, deviates from the overall galactic abundance gradient, having a significantly lower oxygen abundance, 12+log(O/H)$_{Te}$$=$7.96$\,\pm\,$0.08, compared to  nearby objects.
One additional neighboring nebula, H128, has instead a large metallicity value,
12+log(O/H)$_{Te}$$=$8.45$\,\pm\,$0.04, for its galactocentric distance, as already found by
\citet{Kennicutt:2003}. 
These results provide  evidence that marginally significant deviations from local metallicity homogeneity can
arise in the interstellar medium of this part of the galaxy over spatial scales of a few kpc.

Among the strong-line abundance diagnostics we considered to derive the O/H abundance ratio for the full sample of 79 \hii\ regions we found that the ONS method (\citealt{Pilyugin:2010})
provides the best agreement with the direct method. 
With the ONS method we did not find any significant difference in the abundance scatter between \hii\ regions in the western arc of M101 and the rest of the galaxy, being approximately 0.10 dex in both cases.
The rather large deviations from the main abundance radial gradient measured for H27 and H128 from the direct method are not confirmed by the ONS method. On the other hand, the ONS method provides a large (0.4~dex) spread in the abundance of individual knots in the supergiant \hii\ region NGC~5471, which is not seen in the 
$T_e$-based data, and which 
we interpret as an example of the fact that strong-line method results should always be taken with care, and be considered valid in a `statistical' sense. For individual measurements the $T_e$ method should be preferred. On the other hand, the result of our test of possible large-scale azimuthal abundance trends between two opposite sides of M101 using strong-line methods, i.e.~that there is no detectable variation,
should be considered robust. The small scatter we generally observe in the abundance gradient, when we consider that it includes \hii\ regions on opposite sides of the galaxy, is also a good indication for a fairly homogeneous azimuthal abundance distribution.

In general, findings of large {\it local} deviations from abundance homogeneity obtained from strong-line methods should be verified with direct, $T_e$-based measurements.
We  point out that the abundances of individual emission knots within large ($\sim$1~kpc) \hii\ region complexes are consistent with chemical abundance homogeneity when using $T_e$-based data.
Besides the case of NGC~5471 mentioned above, this is true for the  NGC~5447 (H128, H143, H149) and
NGC~5462 (H1159, H1170, H1176) complexes. On the other hand, claiming sizable inhomogeneities from abundances measured from some strong-line indicators appears risky. For example, the N2 parameter (in both the \citealt{Pettini:2004} and \citealt{Perez-Montero:2009a} versions of the calibration considered here) yields abundance variations on the order of $\sim$0.2 dex between the few knots contained in the \hii\ region complexes mentioned above, i.e.~over spatial scales of a few hundred pc. Among the N-based diagnostics, the NO indicator calibrated by \citet{Bresolin:2007} appears much more robust, yielding variations on the order of 0.03 dex among these objects.

It is tempting to try and interpret the relatively small abundance peculiarity we detected in the western disk of M101 using  the direct method with the notion that metallicity inhomogeneities could arise from
metal-poor gas infall and galaxy interactions. The presence of high-velocity clouds and distortions in 
the spiral structure in the eastern disk of M101 \citep{van-der-Hulst:1988,Waller:1997,Sancisi:2008}
can be attributed to tidal interactions between M101 and one or more of its surrounding neighboring dwarf galaxies. Very recently, \citet{Mihos:2012} presented a new deep \hi\ map of the M101 group, which revealed
the presence of newly-discovered  \hi\ clouds and an \hi\ `plume' extending from the SW of M101, likely of tidal nature. It is of course difficult to associate any of these features with the observed metallicity patterns in the M101 disk, but it is tantalizing to observe that the marginally larger scatter in the abundance distribution that we detect in the western arc is taking place  on the same side of the galaxy where these tidal features appear now more evident. Future studies should address this possibility by 
obtaining a large number of high-quality spectroscopic observations and direct abundances in this part of M101.

\vspace*{0.2cm}
\acknowledgments
YL would like to thank J.~Patrick Henry for statistical consult about the errors on the scatters, R.P. Kudritzki for useful comments, and T.-T. Yuan for her help with various scientific discussions.
FB gratefully acknowledges partial support from the National Science Foundation grants AST-0707911 and AST-1008798. We thank the anonymous referee for constructive comments that helped us to improve the quality of the paper.

\clearpage

\bibliography{Papers}

\end{document}